\documentclass[a4paper,11pt]{article}

\usepackage{jcappub} 

\usepackage[T1]{fontenc} 
\usepackage{graphicx}
\usepackage{dcolumn}
\usepackage{bm}
\usepackage{amsmath}
\usepackage{amssymb}
\usepackage{hyperref}

\makeatletter

\title{\boldmath Continuous Sensitivity Analysis for $\delta N$ Formalism}

\author[a]{S. Mohammad Ahmadi}

\affiliation[a]{Department of Physics, University of Tehran, Karegar Ave North, Tehran 14395-547, Iran}

\emailAdd{mohammadahmadi@ut.ac.ir}

\abstract{
The $\delta N$ formalism provides a powerful non-perturbative framework for following the evolution of primordial curvature perturbations on super-horizon scales. However, its standard implementation relies on the separate universe assumption, which neglects significant spatial gradient interactions. Recent work has addressed this limitation by incorporating gradient interactions directly into the background dynamics through an effective source term in the Klein--Gordon equation, thereby extending the applicability of the $\delta N$ framework beyond the separate universe approximation. Despite this conceptual progress, practical calculations within the $\delta N$ formalism remain technically challenging, as cosmological observables require evaluating the sensitivity of the total number of $e$-folds to initial conditions, a task that becomes even more involved once gradient contributions are included. In this work, we develop a systematic method to simplify these calculations by applying Continuous Sensitivity Analysis to the gradient-corrected $\delta N$ framework. In this approach, the required phase-space derivatives are obtained by solving a set of coupled first-order differential equations for the field Jacobian and Hessian, which significantly streamlines both analytical and numerical evaluations of the $\delta N$ formalism. As an explicit demonstration, we apply the method to the Starobinsky model, which features a sharp transition into an ultra-slow-roll phase. Within this setup, we derive analytical expressions for the $k$-dependent power spectrum including full gradient corrections, and obtain an analytical estimate of the equilateral non-Gaussianity parameter $f_{\rm NL}^{\rm eq}$ that accurately captures the main non-Gaussian features.
}


\arxivnumber{2603.27366} 

\begin{document}
\maketitle
\flushbottom

\section{Introduction}

The theory of cosmic inflation provides a highly successful paradigm for the early universe, naturally explaining its large-scale homogeneity and isotropy while generating the primordial curvature perturbations that seed structure formation \cite{starobinsky1980new, sato1981first, guth1981inflationary, linde1982new, albrecht1982reheating, lyth1999particle}. An accurate prediction of the evolution of these curvature perturbations is of paramount importance, not only for understanding the cosmic microwave background and large-scale structure, but also for predicting the correct abundance of primordial black holes (PBHs) \cite{zel1966hypothesis, hawking1971gravitationally, carr1974black, carr1975primordial, chapline1975cosmological, carr1984can}. Formed from the collapse of large density fluctuations in the early universe, PBHs have recently garnered significant attention as a promising candidate for dark matter \cite{green2021primordial, cole2023primordial}. 

To track the nonlinear evolution of these perturbations, the $\delta N$ formalism has emerged as a powerful theoretical tool \cite{starobinsky1982dynamics, starobinskiǐ1985multicomponent, sasaki1996general, sasaki1998super, lyth2005general, pattison2019stochastic, artigas2022hamiltonian}. Its primary advantage lies in its non-perturbative nature, which allows for the evaluation of cosmological observables on super-horizon scales without the need to solve the full perturbed field equations. However, standard implementations of the $\delta N$ formalism strictly rely on the separate universe assumption \cite{wands2000new, lyth2003conserved, rigopoulos2003separate}, which assumes that each local patch of the universe evolves like an unperturbed Friedmann-Lema\^{i}tre-Robertson-Walker (FLRW) spacetime, thereby neglecting spatial gradient interactions. It has been shown that this formalism breaks down at certain scales \cite{jackson2024separate, briaud2025stochastic}, particularly in models featuring transient phases of ultra-slow-roll (USR) evolution \cite{tsamis2004improved, kinney2005horizon, namjoo2013violation, martin2013ultra}. Consequently, the formalism can fail to accurately capture nonlinear effects and non-Gaussianities, whose precise characterization is crucial for reliable predictions of the PBH abundance \cite{byrnes2012primordial, young2013primordial, passaglia2019primordial, atal2019role, biagetti2021formation, taoso2021non, young2022peaks, ferrante2022primordial, gow2023non}.

Recently, significant progress has been made to incorporate spatial gradients into the $\delta N$ framework. For instance, the authors of Ref.~\cite{artigas2025extended} demonstrated that the leading adiabatic gradient correction, $\mathcal{O}(k^2)$, can be effectively encoded in the spatial curvature of each locally homogeneous FLRW slice. Alternatively, a more systematic approach was introduced in Ref.~\cite{ahmadi2026delta}, which successfully encodes full gradient interactions directly into the background Klein--Gordon (KG) equation through an effective source term. This method enables the precise tracking of the linear evolution of curvature perturbations while retaining the non-perturbative advantages of the $\delta N$ formalism to analyze complex nonlinear effects. Another alternative approach to bypass the separate universe assumption is the higher-order matching method~\cite{jackson2024separate}. However, evaluating $k^2$ corrections within this matching framework remains exceedingly difficult; to the best of our knowledge, such evaluations have thus far only been performed numerically \cite{jackson2024separate, ahmadi2026delta}. Moreover, the matching method is strictly linear in perturbation theory and therefore cannot be used to study deviations from Gaussianity.

Despite its theoretical appeal, $\delta N$ calculations are generally challenging. Evaluating observables requires finding the number of $e$-folds, as an explicit function of initial conditions, and subsequently computing its phase-space derivatives. Analytical evaluations within the standard $\delta N$ formalism are consequently limited to a small class of models. The inclusion of gradient interaction terms—either as a source term in the KG equation, as proposed in \cite{ahmadi2026delta}, or through modifications of the spatial curvature, as discussed in \cite{artigas2025extended}—renders these evaluations even more formidable. Even in related applications, such as using the $\delta N$ formalism to directly evaluate the probability density function (PDF) of the curvature perturbation---where scale locality and scale invariance are usually assumed to make calculations more tractable---the inclusion of gradients remains elusive \cite{wands2010local, ezquiaga2020exponential, hooshangi2022rare, cai2022highly, hooshangi2023tail, pi2023logarithmic, kawaguchi2023highly, ballesteros2024intrinsic, ballesteros2024non, cruces2025delta}. While finding $\delta N$ as a function of field perturbations is usually easier than evaluating the power spectrum or non-Gaussianity, recent attempts to facilitate these PDF calculations still neglect gradient interactions \cite{cruces2025delta}. This work focuses on the evaluation of the power spectrum and the leading non-Gaussian effects, rather than the PDF.

These observations highlight a significant drawback of the $\delta N$ formalism: its gradient-corrected predictions are exceedingly difficult to follow analytically. In this work, we address this challenge by introducing Continuous Sensitivity Analysis (CSA) \cite{cacuci1981sensitivity1, cacuci1981sensitivity2, griewank2008evaluating, kim2005efficient, cacuci2005sensitivity, banks2012estimation} to the $\delta N$ formalism, a mathematical framework that drastically simplifies the evaluation of linear and nonlinear $\delta N$ observables. CSA is a rigorous technique used to continuously track how the final state of a dynamical system responds to small variations in its parameters or initial conditions by deriving and solving a set of coupled differential equations for the sensitivities themselves. This method is exceptionally well-suited to the $\delta N$ framework, as the formalism is inherently defined by the sensitivity of the final $e$-folding number to initial field values.

Related ideas have appeared previously in the literature. In particular, Ref.~\cite{yokoyama2007SR, yokoyama2008primordial} considered deriving evolution equations for quantities entering the $\delta N$ expansion by differentiating the background field equations. However, our approach differs in both formulation and scope. In the present work, we develop the analysis from first principles within the framework of CSA, deriving explicit evolution equations for the Jacobian and Hessian quantities that determine the $\delta N$ coefficients. This framework is particularly well suited to our aims: incorporating gradient corrections into the formalism and comparing the resulting exact expressions with those obtained in linear perturbation theory and in the separate-universe approach.

We discuss the application of CSA to evaluate both linear and second-order $\delta N$ outputs. As a concrete example, we apply this method to the Starobinsky model \cite{starobinskij1992spectrum, martin2012scalar, martin2014sharp, ahmadi2022quantum}. We employ CSA to find the homogeneous solution relevant to the standard $\delta N$ formalism and use the Green's function method to evaluate the gradient corrections, particularly the leading-order $k^2$ corrections. We then introduce a simple identification that allows for the analytical evaluation of $\delta N$ with full gradient corrections. Proceeding to the second-order sensitivity analysis, we find the non-Gaussianity with partial gradient inclusions. We demonstrate that our analytical approach closely matches results derived from full numerical methods.

Beyond its analytical utility, the CSA framework offers some theoretical advantages. Traditionally, proving the equivalence between the $\delta N$ formalism and linear perturbation theory relies on comparing the equations of motion for field perturbations \cite{pattison2019stochastic, ahmadi2026delta}. The CSA, however, allows for a direct comparison between the final output of the $\delta N$ formalism and the evolution of curvature perturbations derived from linear perturbation theory. This provides a more stringent method for checking the validity of the $\delta N$ approach, enabling us to precisely identify regimes where the formalism breaks down. Using this technique, we rigorously demonstrate that the $\delta N$ formalism fails in regimes characterized by a large Hubble flow parameter, $\epsilon$. This failure occurs because the momentum constraint is missing in the exact separate universe picture \cite{kodama1998evolution, sasaki1998super}, and the source term for gradient interactions introduced in \cite{ahmadi2026delta} is not designed to retain the effect of this missing constraint. By comparing the first-order sensitivity equations with the perturbation equations, we introduce a corrected source term designed to capture the subtle interaction effects arising from the momentum constraint. Through full numerical evaluations, we demonstrate that this corrected source term successfully reproduces the results of linear perturbation theory in scenarios featuring transient large-$\epsilon$ regimes, where standard implementations of the $\delta N$ formalism break down.

 Finally, the CSA method is highly advantageous for numerical evaluations. By reducing the full second-order evolution of curvature perturbations to a system of first-order differential equations, it renders numerical computations significantly faster, more stable, and more accurate.

This paper is organized as follows. In Section \ref{review_dN_fullgrad}, we provide a brief review of the $\delta N$ formalism with gradient interactions as introduced in Ref.~\cite{ahmadi2026delta}. In Section \ref{CSA_Formulation}, we construct the linear and second-order CSA frameworks. We derive the sensitivity equations necessary to evaluate the power spectrum and non-Gaussianity parameters, and prove the equivalence of the CSA method to linear perturbation theory (including its breakdown at large-$\epsilon$ regimes and how it can be fixed). In Section \ref{sec_example_staro}, we apply the CSA methodology to the Starobinsky model, analytically finding the power spectrum and equilateral non-Gaussianity parameter with gradient interactions, and comparing them against full numerical results. Finally, the Appendices provide the evaluation of specific relations and illustrations of specific technical aspects of the CSA framework. Throughout the paper, we work in Planck units where $ c = G = \hbar = M_{\mathrm{Pl}} = 1 $.

\section{\boldmath Review of Gradient Expansion and $\delta N$ Formalism}
\label{review_dN_fullgrad}

The evolution of cosmological perturbations on superhorizon scales is commonly studied using long-wavelength techniques that exploit the smallness of spatial gradients. Two widely used frameworks in this context are the gradient expansion method and the $\delta N$ formalism \cite{starobinsky1982dynamics, starobinskiǐ1985multicomponent, sasaki1996general, sasaki1998super, lyth2005general, pattison2019stochastic, artigas2022hamiltonian}. The gradient expansion provides a systematic perturbative treatment of spatial gradient effects by expanding the solution of the Mukhanov--Sasaki equation in powers of $k^2$, while the $\delta N$ formalism is based on the separate universe approach (SUA) \cite{wands2000new, lyth2003conserved, rigopoulos2003separate}, which relates the curvature perturbation to fluctuations in the local expansion history. In this section, we briefly review the essential aspects of both approaches and the matching procedures used to connect subhorizon quantum fluctuations with their superhorizon evolution. This discussion will also clarify how gradient corrections can be incorporated into the $\delta N$ framework and how the resulting formulation relates to the gradient expansion.

\subsection{Gradient Expansion and Matching Methods}

The evolution of scalar perturbations during inflation is conveniently described by the comoving curvature perturbation $\mathcal{R}_k$. At linear order, its dynamics are governed by the Mukhanov--Sasaki (MS) equation
\begin{equation}
	\mathcal{R}_k'' + 2\frac{z'}{z}\mathcal{R}_k' + k^2 \mathcal{R}_k = 0 ,
	\label{MS_equation_R}
\end{equation}
where primes denote derivatives with respect to conformal time $\tau$, and the quantity $z \equiv a\dot{\phi}/H$ encodes the background evolution of the inflaton field. For future convenience, we also define the hierarchy of Hubble flow parameters. The first three parameters are explicitly given by
\begin{equation}
	\epsilon \equiv - \frac{d \ln H}{d n}, \qquad
	\eta \equiv \frac{d \ln \epsilon}{d n}, \qquad
	\xi \equiv \frac{d \ln \eta}{d n},
	\label{eq_hubble_flow_params}
\end{equation}
where $n$ is the number of $e$-folds defined by $n \equiv \ln a$. To facilitate comparison between our later $\delta N$ calculation and the MS equation, we also rewrite equation \eqref{MS_equation_R} in terms of the $e$-fold number as
\begin{equation}
	\frac{d^2 \mathcal{R}_k}{dn^2} + ( 3 - \epsilon + \eta ) \frac{d \mathcal{R}_k}{dn} + \frac{k^2}{a^2 H^2} \mathcal{R}_k = 0.
	\label{exact_MS_N}
\end{equation}

In the long-wavelength limit $k \rightarrow 0$, the gradient term becomes negligible and the equation admits the homogeneous solution
\begin{equation}
	\mathcal{R}_h(\tau)
	=
	C_k
	+
	D_k
	\int_{\tau}^{\tau_{\rm ref}}\!\!\!
	\frac{d\tilde{\tau}}{z^2(\tilde{\tau})} ,
	\label{homogeneous_solution_MS}
\end{equation}
where $C_k$ and $D_k$ are integration constants determined by initial conditions and $\tau_{\rm ref}$ is an arbitrary reference time. The first term corresponds to the adiabatic growing mode, while the second represents a decaying mode.

To systematically incorporate the effects of spatial gradients, one may construct a gradient expansion in powers of $k^2$. Writing the curvature perturbation as the sum of growing and decaying modes, one can expand
\begin{equation}
	\mathcal{R}_k(\tau)
	=
	\sum_{i=0}^{\infty} \mathcal{G}_i(\tau) k^{2i}
	+
	\sum_{i=0}^{\infty} \mathcal{D}_i(\tau) k^{2i},
	\label{expansions_GD}
\end{equation}
where the leading-order terms reproduce the homogeneous solution \eqref{homogeneous_solution_MS}. That means
\begin{equation}
	\mathcal{G}_0 = C_k, \qquad \mathcal{D}_0 = D_k \int_{\tau}^{\tau_{\rm ref}}\!\!\!
	\frac{d\tilde{\tau}}{z^2(\tilde{\tau})}.
\end{equation}
Higher-order corrections are generated through recursive integral relations
\begin{equation}
	\mathcal{D}_i(\tau)
	= -
	\int_{\tau_{\rm ref}}^{\tau}\!\!
	\frac{d\tau_1}{z^2(\tau_1)}
	\int_{\tau_{\rm ref}}^{\tau_1}\!
	d\tau_2 \,
	z^2(\tau_2)\,
	\mathcal{D}_{i-1}(\tau_2),
	\label{Di_parameters}
\end{equation}
with an analogous expression for $\mathcal{G}_i$. In principle, the full solution of the MS equation is recovered by summing the complete series, which systematically incorporates all gradient interactions.

In practice, the connection between the quantum generation of perturbations and their classical superhorizon evolution is implemented through a matching procedure \cite{jackson2024separate}. In the standard \emph{homogeneous matching} approach, the full solution of the MS equation is matched at a time $\tau_*$, defined by $k = \sigma a(\tau_*) H(\tau_*)$, to the homogeneous solution. Requiring continuity of $\mathcal{R}_k$ and $\mathcal{R}'_k$ determines the integration constants,
\begin{equation}
	\hat{C}_k
	=
	\mathcal{R}_{k*}
	+
	z_*^2 \mathcal{R}'_{k*}
	\int_{\tau_*}^{0}\!
	\frac{d\tilde{\tau}}{z^2(\tilde{\tau})},
	\qquad
	\hat{D}_k
	=
	- z_*^2 \mathcal{R}'_{k*},
	\label{homoheneous_hatCD}
\end{equation}
where starred quantities are evaluated at the matching time, and hatted quantities denote values obtained via the matching procedure.

When gradient effects remain relevant after horizon exit, the homogeneous approximation becomes insufficient. A systematic improvement is provided by the \emph{higher-order matching} method, in which the superhorizon solution is written as
\begin{equation}
	\hat{\mathcal{R}}_k(\tau)
	=
	\mathcal{R}_{k*} u_{\rm ad}(\tau)
	+
	\mathcal{R}'_{k*} u_{\rm nad}(\tau),
	\label{R_higherorder_Matching}
\end{equation}
where the mode functions $u_{\rm ad}$ and $u_{\rm nad}$ are determined by matching the higher-order terms of the gradient expansion (derived by relation~\eqref{Di_parameters}) to the MS solution. This procedure incorporates finite-$k$ corrections and enables one to track the evolution of perturbations beyond the leading long-wavelength approximation.

\subsection{$\delta N$ Formalism With Gradient Interactions}
\label{sec_dn_with_gradients}

The SUA provides an intuitive description of superhorizon perturbations. In the presence of scalar perturbations, inhomogeneities enter the metric through spatial gradient operators $\partial_i$. On sufficiently large scales, where $k \rightarrow 0$ and gradient terms can be neglected, each local Hubble patch evolves approximately as a homogeneous and isotropic universe with slightly perturbed background parameters. In this limit, the dynamics of the scalar field in each patch are governed by the homogeneous KG equation
\begin{equation}
	\ddot{\phi}+3H\dot{\phi}+V_{,\phi}=0,
	\label{KG_homo}
\end{equation}
where a dot denotes differentiation with respect to cosmic time $t$, and the comma subscript indicates a derivative. The key idea is that long-wavelength perturbations can be absorbed into locally perturbed background quantities, allowing the evolution of each patch to be treated as an independent FLRW universe.

To connect this picture with linear perturbation theory, it is convenient to introduce the MS variable
\begin{equation}
	Q_k = \frac{\dot{\phi}}{H}\mathcal{R}_k .
\end{equation}
Substituting this relation into the MS equation yields the evolution equation \cite{jackson2024separate, ahmadi2026delta}
\begin{equation}
	\ddot{Q}_k + 3H\dot{Q}_k
	+
	\left(
	\frac{k^2}{a^2}
	+
	V_{,\phi\phi}
	-
	\frac{1}{a^3}
	\frac{d}{dt}\left(\frac{a^3}{H}\dot{\phi}^2\right)
	\right)Q_k
	=0 ,
	\label{perturbed_KG}
\end{equation}
which describes the full linear dynamics of scalar perturbations. On the other hand, the perturbed background equation obtained by linearizing the homogeneous dynamics \eqref{KG_homo} in the spatially flat or uniform-expansion gauge can be written as
\begin{equation}
	\ddot{Q}_k + 3H\dot{Q}_k
	+
	\left(
	V_{,\phi\phi}
	-
	\frac{1}{a^3}
	\frac{d}{dt}\left(\frac{a^3}{H}\dot{\phi}^2\right)
	\right)Q_k
	=0 .
	\label{perturbed_BG}
\end{equation}
Comparing equations \eqref{perturbed_KG} and \eqref{perturbed_BG} shows that the separate universe description corresponds precisely to the $k \rightarrow 0$ limit of the full perturbation dynamics.

The $\delta N$ formalism provides a practical realization of the SUA. It states that on superhorizon scales the curvature perturbation can be identified with the perturbation in the local expansion history, $\zeta \simeq \delta N$. In this framework, the curvature perturbation is obtained by evolving locally homogeneous background trajectories with slightly perturbed initial conditions,
\begin{equation}
	\zeta_k
	=
	N(\phi_*+\delta\phi_{k*},\dot{\phi}_*+\delta\dot{\phi}_{k*})
	-
	N(\phi_*,\dot{\phi}_*).
\end{equation}
Here $N(\phi_*,\dot{\phi}_*)$ denotes the number of $e$-folds realized between the beginning and the end of inflation for the initial conditions $\phi_*$ and $\dot{\phi}_*$. Expanding in the initial perturbations yields
\begin{equation}
	\delta N_k
	=
	N_a\,\delta X^a_{k}
	+
	\frac12
	N_{ab}
	\int \!
	\frac{d^3q}{(2\pi)^3}
	\delta X^a_{q}\,
	\delta X^b_{k-q}
	+\cdots ,
	\label{dN_expansion_standard}
\end{equation}
where $X=(\phi_*,\dot{\phi}_*)$ represents the set of initial conditions and derivatives of $N$ are defined by
\begin{equation}
	N_a \equiv \left. \frac{\partial N}{\partial X^a} \right|_{\tau_*},
	\qquad
	N_{ab} \equiv \left. \frac{\partial^2 N}{\partial X^a \partial X^b} \right|_{\tau_*}.
\end{equation}

The strength of the $\delta N$ formalism lies in its ability to capture nonlinear effects and non-Gaussian statistics. However, it relies on the SUA and therefore neglects spatial gradients, which may become important in non-attractor phases. To incorporate gradient effects within the $\delta N$ framework, one may modify the background KG equation by introducing an effective source term that captures the missing spatial Laplacian contribution. The corrected background equations take the form \cite{ahmadi2026delta}
\begin{subequations}\label{BG_equations_corrected}
	\begin{align}
		3 H^2 - \frac{\dot{\phi}^2}{2} - V =&\, 0, \\
		\ddot{\phi}+3H\dot{\phi}+V_{,\phi} =&\, \mathcal{S}_k , \label{KG_with_source}
	\end{align}
\end{subequations}
where the source term is chosen as
\begin{equation}
	\mathcal{S}_k = -\frac{k^2\dot{\phi}}{a^2H}\,\hat{\mathcal{R}}_k .
	\label{source_term}
\end{equation}
The origin of this correction term can be understood by comparing the perturbed background equation \eqref{perturbed_BG} with the full perturbation equation \eqref{perturbed_KG}. The discrepancy arises from the missing gradient contribution $k^2Q_k/a^2$ in the SUA. Using the relation $Q_k=\dot{\phi}\mathcal{R}_k/H$, this term can be reinterpreted as an effective source acting on the background field evolution.

One can further demonstrate that this source-term formulation is formally equivalent to the gradient-expansion solution. Linearizing the modified equation \eqref{KG_with_source} and substituting $Q_k = z\mathcal{R}_k/a$ leads to
\begin{equation}
	\left(z^2\mathcal{R}_k'\right)' = -k^2 z^2 \hat{\mathcal{R}}_k .
\end{equation}
Expanding $\mathcal{R}_k$ in powers of $k^2$ and isolating the contribution at order $k^{2(i+1)}$ yields
\begin{equation}
	\left(z^2 \mathcal{R}_k^{(i+1)\prime}\right)'
	=
	- z^2 \left(\mathcal{G}_i+\mathcal{D}_i\right).
\end{equation}
Integrating twice gives
\begin{equation}
	\mathcal{R}_k^{(i+1)}(\tau)
	=-
	\int_{\tau_*}^{\tau}\! \frac{d\tau_1}{z^2(\tau_1)}
	\int_{\tau_*}^{\tau_1} \!
	d\tau_2\, z^2(\tau_2)
	\left(\mathcal{G}_i(\tau_2)+\mathcal{D}_i(\tau_2)\right)
	=
	\mathcal{G}_{i+1}(\tau)+\mathcal{D}_{i+1}(\tau).
	\label{R_equal_gradient}
\end{equation}
The right-hand side corresponds to the $(i+1)$th term in the gradient expansion \eqref{expansions_GD}. This establishes the equivalence between the source-term formulation and the higher-order matching method. Consequently, the $\delta N$ formalism can be extended beyond the strict long-wavelength approximation by incorporating this effective source term, allowing the nonlinear advantages of the $\delta N$ approach to be retained while consistently accounting for gradient interactions.

Having established the correspondence between the gradient-corrected $\delta N$ framework and the gradient-expansion solution at linear order, one can extend the analysis to higher-order perturbations in order to quantify primordial non-Gaussianity \cite{komatsu2001acoustic, maldacena2003non, bartolo2004non, liguori2010primordial, chen2010primordial, wang2014inflation, cai2018revisiting, namjoo2025geometry}. A convenient parametrization of the leading non-Gaussian signal is provided by the nonlinear parameter $f_{\rm NL}$, which characterizes the amplitude of the curvature perturbation bispectrum. By utilizing the $\delta N$ expansion \eqref{dN_expansion_standard} up to second order, one can explicitly compute the shape function of the three-point correlator as
\begin{equation}
	S(\mathbf{k}_1,\mathbf{k}_2,\mathbf{k}_3)
	=
	\frac{5}{6}
	\frac{k_3^3 N_a N_b N_{cd}\,
		\mathcal{P}^{ac}(k_1)\mathcal{P}^{bd}(k_2)
		+\text{2 perms}}
	{k_1 k_2 k_3}.
\end{equation}
Here $\mathcal{P}^{ab}(k)$ denotes the cross-power spectrum of the initial phase-space perturbations, defined by
\begin{equation}
	\mathcal{P}^{ab}(k)
	\equiv
	\frac{k^3}{2\pi^2}
	\mathrm{Re}
	\left[
	\delta X^a_k\, (\delta X^{b}_k)^*
	\right].
	\label{power_spectra_definition}
\end{equation}
Comparing this with the standard definition of the shape function,
\begin{equation}
	S(\mathbf{k}_1,\mathbf{k}_2,\mathbf{k}_3)
	=
	f_{\rm NL}\,
	\frac{k_3^3 \mathcal{P}_{\mathcal{R}}(k_1)\mathcal{P}_{\mathcal{R}}(k_2)+\text{2 perms}}
	{k_1 k_2 k_3},
\end{equation}
one finds the following relation for $f_{\rm NL}$:
\begin{equation}
	f_{\rm NL}
	=
	\frac{5}{6}
	\frac{k_3^3 N_a N_b N_{cd}\,
		\mathcal{P}^{ac}(k_1)\mathcal{P}^{bd}(k_2)
		+\text{2 perms}}
	{k_3^3 N_e N_f N_g N_h\,
		\mathcal{P}^{ef}(k_1)\mathcal{P}^{gh}(k_2)
		+\text{2 perms}} .
		\label{non_gaussian_fNL}
\end{equation}
This relation provides a convenient way to compute the non-Gaussianity once the derivatives of $N$ and the statistics of the initial perturbations are known.

\section{\boldmath Continuous Sensitivity Analysis in $\delta N$ Formalism}
\label{CSA_Formulation}

In this section, we discuss Continuous Sensitivity Analysis (CSA) \cite{cacuci1981sensitivity1, cacuci1981sensitivity2, griewank2008evaluating, kim2005efficient, cacuci2005sensitivity, banks2012estimation} as a systematic mathematical framework for the analytical evaluations of the $\delta N$ formalism. While the traditional $\delta N$ approach relies on determining the final number of $e$-folds as an explicit function of initial conditions—a task that is often analytically challenging when spatial gradient interactions are included—CSA shifts the focus to tracking the continuous dynamical evolution of the system's sensitivity to these initial perturbations through a set of coupled first-order differential equations. Furthermore, this formulation can naturally accommodates the effective source term of gradient interactions introduced in Section \ref{sec_dn_with_gradients}.

\subsection{Linear CSA Formulation}

We define the state vector consisting of the background field and its number of $e$-fold derivative as
\begin{equation}
	\left[Y^i\right] = \begin{pmatrix} \phi \\ \Pi \end{pmatrix},
\end{equation}
where $\Pi \equiv \phi_{,n}$. From this point onward, square brackets will be used to denote the vector or matrix representation of the components $Y^i$.

Let $n_e$ denote the time at the end of inflation and let $X = (\phi_*, \Pi_*)$ represent the set of initial conditions defined on the matching hypersurface. We assume that the final hypersurface corresponds to the comoving gauge, in which the field fluctuation vanishes, $\delta\phi_e=0$. Varying $\phi_e = \phi(X,n_e)$ and keeping first-order terms then yields (see Appendix \ref{dN_expansion_appendix})
\begin{equation}
	\delta N_k = - \left. \left[\frac{1}{\phi_{,n}}  \frac{\partial \phi}{\partial X^a} \right] \right|_{n_e} \delta X^a_k =  -\frac{1}{\Pi_e} J^1_a \, \delta X^a_k ,
	\label{linear_dN_expansion_tau}
\end{equation}
where $J^i_a \equiv \partial  Y^i/ \partial X^a$ is the sensitivity matrix (Jacobian) that measures the dependence of the field values on the final hypersurface with respect to variations in the initial conditions $X$. From now on, we drop the subscript $k$ from all parameters for the sake of simplicity.

Therefore, obtaining the linear $\delta N$ result reduces to computing the elements $J^i_a$. To this end, we derive the differential equation governing the evolution of $J$. Writing the gradient corrected background equations \eqref{BG_equations_corrected} in terms of $e$-fold number and removing the Hubble parameter between them yields
\begin{equation}
	\phi_{,nn} = \left(\epsilon - 3\right) \left(g + \Pi \right)  + \left( \frac{\epsilon - 3}{V} \right) \frac{k^2}{a^2} \Pi \hat{\mathcal{R}},
\end{equation}
where $g \equiv V_{,\phi} / V$. This can be used to write the derivative of the state vector in the form
\begin{equation}
	(Y^i)_{,n} = F^i + \Lambda^i ,
	\label{BG_evolve_tau}
\end{equation}
where
\begin{equation}
	\left[{F}^i\right] = \begin{pmatrix} \Pi \\ (\epsilon - 3)(g + \Pi) \end{pmatrix}, \qquad 
	{\Lambda}^i = {B} {\mathcal{W}}^i,
	\label{definition_F_Lambda}
\end{equation}
with
\begin{equation}
	{B} = \left( \frac{\epsilon - 3}{V} \right) \frac{k^2}{a^2} \Pi = -\frac{k^2}{a^2 H^2} \Pi
	\label{definition_B}, \qquad \left[{\mathcal{W}}^i\right] = \begin{pmatrix} 0 \\ \hat{\mathcal{R}}  \end{pmatrix}.
\end{equation}
Here $F^i$ encodes the background dynamics, while $\Lambda^i$ represents the gradient-sourced correction.

Taking the derivative of the equations of motion \eqref{BG_evolve_tau} with respect to the initial conditions $X$ leads to a matrix differential equation governing the Jacobian, which we can the sensitivity equation:
\begin{equation}
	(J^i_a)_{,n} = A^i_j J^j_a + \Sigma^i_a ,
	\label{J_Evolve_tau}
\end{equation}
where the stability matrix is defined by $A^i_j \equiv \partial F^i/\partial Y^j$, and the sensitivity forcing matrix by $\Sigma^i_a \equiv \partial \Lambda^i/\partial X^a$. Since $n_*$ denotes the initial hypersurface on which $Y = X$, the Jacobian initial condition is therefore given by
\begin{equation}
	\left[J_*\right] =
	\begin{pmatrix}
		\frac{\partial \phi_*}{\partial \phi_*} &
		\frac{\partial \phi_*}{\partial \Pi_*} \\
		\frac{\partial \Pi_*}{\partial \phi_*} &
		\frac{\partial \Pi_*}{\partial \Pi_*}
	\end{pmatrix}
	=
	\begin{pmatrix}
		1 & 0 \\
		0 & 1
	\end{pmatrix}.
	\label{initial_conditions_J}
\end{equation}
The stability matrix $A$ follows directly from the background flow $F$,
\begin{equation}
	\left[A^i_j\right] = \left[ \frac{\partial F^i}{\partial Y^j}\right] = 
	\begin{pmatrix} 
		0 & 1 \\ 
		(\epsilon - 3) g_{,\phi} & \Pi(g + \Pi) + (\epsilon - 3) 
	\end{pmatrix}.
	\label{A_matrix_tau}
\end{equation}

For the first component of the forcing matrix, since $\Lambda^1 = 0$, we find $\Sigma^1_a = 0$. For the second component we can write
\begin{equation}
	\Sigma^2_a = \frac{\partial \Lambda^2}{\partial X^a} = B \frac{\partial \mathcal{W}^2}{\partial X^a}
	+ \frac{\partial B}{\partial X^a} \mathcal{W}^2 .
	\label{sigma_exact}
\end{equation}
Since the second term on the right-hand side contains $\hat{\mathcal{R}}$ explicitly, it already contributes at $\mathcal{O}(1)$ to the Jacobian. When the Jacobian is subsequently multiplied by $\delta X$ to obtain the final perturbation, this term produces a contribution of order $\mathcal{O}(2)$. It therefore lies beyond the linear approximation and can be consistently neglected at linear order. Specializing to the spatially flat gauge at the initial hypersurface, the initial state perturbations are related to the curvature perturbation through
\begin{align}
	\delta\phi_* = \Pi_* \mathcal{R}_* , \qquad \delta\Pi_* = \Pi_* \left( \frac{\eta_*}{2} \mathcal{R}_* + \frac{\mathcal{R}'_*}{a_* H} \right).
	\label{spatially_flat_dpdpi}
\end{align}
The quantity $\hat{\mathcal{R}}$ can be decomposed into adiabatic and non-adiabatic components according to the relation \eqref{R_higherorder_Matching}, allowing it to be expressed as a linear combination of the initial perturbations,
\begin{equation}
	\hat{\mathcal{R}} = C_a \, \delta X^a = C_1 \, \delta\phi_* + C_2 \, \delta\Pi_* ,
\end{equation}
with coefficients mapping to the initial state space defined by
\begin{equation}
	[C_a] =
	\frac{1}{\Pi_*}
	\begin{pmatrix}
		u_{\rm ad} - \frac{1}{2} a_* H \eta_* u_{\rm nad} \\
		a_* H u_{\rm nad}
	\end{pmatrix}.
\end{equation}
Using these, the forcing matrix evaluates to
\begin{equation}
	\left[\Sigma^i_a\right] =
	{B} \begin{pmatrix} 0 & 0 \\ {C}_1 & {C}_2 \end{pmatrix}.
	\label{forcing_matrix_tau}
\end{equation}

Having determined all components entering equation \eqref{J_Evolve_tau}, the explicit form of the first-order sensitivity equations can be written as
\begin{subequations}
	\begin{align}
		(J^1_1)_{,n} &= J^2_1, \\
		(J^1_2)_{,n} &= J^2_2, \\
		({J}^2_1)_{,n} &= (\epsilon - 3)g_{,\phi} {J}^1_1 + \Big[ \Pi(g + \Pi) + (\epsilon - 3) \Big] {J}^2_1 + {B} {C}_1 , \\
		({J}^2_2)_{,n} &= (\epsilon - 3)g_{,\phi} {J}^1_2 + \Big[ \Pi(g + \Pi) + (\epsilon - 3) \Big] {J}^2_2 + {B} {C}_2 .
	\end{align}
\end{subequations}
Substituting the first two equations into the last two yields
\begin{equation}
	({J}^1_a)_{,nn} - \Big[ \Pi(g + \Pi) + (\epsilon - 3) \Big] ({J}^1_a)_{,n} + (3 - \epsilon )g_{,\phi} {J}^1_a = {B} {C}_a .
	\label{SO_J_evolution}
\end{equation}
We are therefore left with a set of decoupled second-order ordinary differential equations that must be solved in order to determine the linear $\delta N$ result.

In the standard implementation of the $\delta N$ formalism using the relation \eqref{dN_expansion_standard}, one must find the inverse of the solution to the KG equation and subsequently differentiate it with respect to the initial conditions in order to obtain the coefficients $N_a$. In contrast, within the CSA framework the Jacobians appearing in the $\delta N$ expansion are obtained directly by solving the second-order differential equations \eqref{SO_J_evolution}. This significantly simplifies the calculation compared with the conventional approach. In particular, this formulation enables one to obtain analytic expressions for the gradient-corrected power spectrum in the Starobinsky model, as demonstrated in Section \ref{FO_CSA_Staro}.

Notice that the gradient-corrected $\delta N$ formalism depends on four initial quantities, 
$\{\phi_*, \Pi_*, \mathcal{R}_*, \mathcal{R}'_*\}$, of which only two are independent. In the above discussion we chose $X = (\phi_*, \Pi_*)$ as the set of independent initial variables. However, alternative choices are possible and can simplify certain aspects of the CSA calculation. Let us instead choose $X = (\mathcal{R}_*, \mathcal{R}'_*)$ and examine how the linear CSA formulation changes. The stability matrix $A^{i}_{j}$ is independent of the choice of $X$, and therefore the relation \eqref{A_matrix_tau} remains valid in this new basis. For the forcing matrix, we obtain an expression analogous to \eqref{forcing_matrix_tau}, but with
\begin{equation}
	[C_{a}] = 
	\begin{pmatrix}
		u_{\mathrm{ad}} \\
		u_{\mathrm{nad}}
	\end{pmatrix}.
	\label{Ci_RRp}
\end{equation}
Consequently, the sensitivity equations \eqref{SO_J_evolution} remain valid in the new basis. Finally, the initial-condition matrix \eqref{initial_conditions_J} becomes
\begin{equation}
	[J_*] = 
	\begin{pmatrix}
		\frac{\partial \phi_*}{\partial \mathcal{R}_*} & \frac{\partial \phi_*}{\partial \mathcal{R}'_*} \\
		\frac{\partial \Pi_*}{\partial \mathcal{R}_*} & \frac{\partial \Pi_*}{\partial \mathcal{R}'_*}
	\end{pmatrix}
	=
	\begin{pmatrix}
		\Pi_* & 0 \\
		\frac{\eta_* \Pi_*}{2} & \frac{\Pi_*}{a_* H}
	\end{pmatrix},
	\label{Js_tau}
\end{equation}
where we have used the relations \eqref{spatially_flat_dpdpi}.

From equation \eqref{Ci_RRp}, it follows that with this choice of initial variables, the evolution of $J^1_1$ depends solely on the adiabatic mode, whereas the evolution of $J^1_2$ involves only the non-adiabatic mode. Throughout this work we adopt this choice of independent initial conditions, as it clarifies several physical aspects of the models and simplifies the analytical calculation of gradient interactions.

\subsection{Second-Order CSA Formulation}
\label{second_order_CSA}

The $\delta N$ expansion, taken to second order in the initial condition perturbations $\delta X^a$, is given by (see Appendix \ref{dN_expansion_appendix} for a detailed derivation):
\begin{equation}
	\delta N = -\frac{1}{\Pi_e} \left[ {J}^1_a \delta X^a + \frac{1}{2} {M}_{ab} \delta X^a \delta X^b \right],
	\label{dN_expansion_second}
\end{equation}
where $M_{ab}$ is defined as
\begin{equation}
	{M}_{ab} \equiv {\Theta}^1_{ab} - \frac{2}{\Pi_e} {J}^2_a {J}^1_b + \frac{\eta_e}{2 \Pi_e} {J}^1_a {J}^1_b.
	\label{M_for_fNL}
\end{equation}
Here, we have defined the second-order sensitivity tensor, or Hessian, as
\begin{equation}
	\Theta^i_{ab} \equiv \frac{\partial^2 Y^i}{\partial X^a \partial X^b}.
\end{equation}
Using the relation \eqref{dN_expansion_second}, one can derive the resulting non-Gaussianity parameter $f_{\rm NL}$, analogous to equation \eqref{non_gaussian_fNL}. For an equilateral configuration,\footnote{We note that within the SUA, where each patch evolves independently, only local-type non-Gaussianities can be evaluated at the level of a single patch. Consequently, when we refer to the ``equilateral'' configuration in this work, we mean the equilateral limit of the local (i.e., squeezed) $f_{\rm NL}$, rather than the parameter associated with the intrinsic equilateral shape of the bispectrum. We follow here the same convention as in Ref.~\cite{hazra2013bingo} (see their equation (10) and the surrounding discussion).} the parameter takes the form
\begin{equation}
	f_{\rm NL}^{\rm eq} = -\frac{5 \Pi_e}{6} \frac{{M}_{ab} {J}^1_c {J}^1_d \mathcal{P}^{ac}(k)\mathcal{P}^{bd}(k)}{({J}^1_e {J}^1_f \mathcal{P}^{ef}(k))^2}.
	\label{fNL_eq_CSA}
\end{equation}
While the evolution of the Jacobian $J^i_a$ is known from first-order theory, a calculation of $f_{\rm NL}$ requires the sensitivity equation for the Hessian $\Theta^i_{ab}$.

This can be obtained by differentiating the first-order sensitivity equation \eqref{J_Evolve_tau} with respect to the initial condition $X^b$, which yields the sensitivity equation for the Hessian:
\begin{equation}
	(\Theta^i_{ab})_{,n}
	= A^i_j\,\Theta^j_{ab}
	+ \mathcal{A}^i_{jk}\,J^j_a\,J^k_b
	+ \Sigma^i_{ab}.
	\label{eq_evolution_Hessian_tau}
\end{equation}
Here, two additional quantities are introduced. The first is the second-order stability tensor,
\begin{equation}
	\mathcal{A}^i_{jk} \equiv \frac{\partial^2 F^i}{\partial Y^j \partial Y^k}
	= \frac{\partial A^i_j}{\partial Y^k},
\end{equation}
and the second is the second-order forcing tensor,
\begin{equation}
	\Sigma^i_{ab} \equiv \frac{\partial^2 \Lambda^i}{\partial X^a \partial X^b}
	= \frac{\partial \Sigma^i_a}{\partial X^b}.
\end{equation}
The evolution of the Hessian begins from the trivial initial condition $\Theta^i_{ab}(n_*) = 0$, since the map $Y(n_*)$ is linear in $X$.

To solve equation \eqref{eq_evolution_Hessian_tau}, we must compute the necessary components of the new tensors. We begin by evaluating ${\mathcal{A}}^i_{jk}$. Since ${F}^1 = \Pi$ is linear in ${Y}$, all of its second derivatives vanish identically (i.e., ${\mathcal{A}}^1_{jk}=0$), which leaves us with
\begin{equation}
	\bigl[{\mathcal{A}}^2_{jk}\bigr] =
	\begin{pmatrix}
		(\epsilon - 3)\,g_{,\phi\phi} & \Pi\,g_{,\phi} \\[4pt]
		\Pi\,g_{,\phi}             & g + 3\Pi
	\end{pmatrix}.
	\label{A2jk}
\end{equation}
For the forcing term, we note that since $\Lambda^1 = 0$, its derivatives also vanish, so $\Sigma^1_{ab} = 0$. The second component, $\Lambda^2 = B\,\hat{\mathcal{R}}$, requires careful differentiation:
\begin{equation}
	\Sigma^2_{ab} = \underbrace{\frac{\partial^2 B}{\partial X^a \partial X^b}\hat{\mathcal{R}}}_{\text{neglected}}
	+ \frac{\partial B}{\partial X^a}\frac{\partial \hat{\mathcal{R}}}{\partial X^b}
	+ \frac{\partial B}{\partial X^b}\frac{\partial \hat{\mathcal{R}}}{\partial X^a}
	+ B\,\underbrace{\frac{\partial^2 \hat{\mathcal{R}}}{\partial X^a \partial X^b}}_{=0}.
\end{equation}
The first term is proportional to $\hat{\mathcal{R}}$ and is therefore neglected at the linear level. The final term vanishes because $\hat{\mathcal{R}}$ is linear in the initial-condition perturbations, and therefore its second derivatives vanish. The surviving terms constitute the second-order forcing tensor:
\begin{align}
	\Sigma^2_{ab}
	= \frac{\partial B}{\partial X^a}\frac{\partial \hat{\mathcal{R}}}{\partial X^b}
	+ \frac{\partial B}{\partial X^b}\frac{\partial \hat{\mathcal{R}}}{\partial X^a} = C_b(B_j J^j_a) + C_a( B_j J^j_b ),
	\label{sigma3_exact}
\end{align}
where we have used the relation $\partial \hat{\mathcal{R}} / \partial X^b = C_b$, and $B_j \equiv \partial B/\partial Y^j$ is given by
\begin{equation}
	{B}_j = \frac{k^2}{a^2 V} \begin{pmatrix} g \Pi (3- \epsilon) \\ -3 (1 - \epsilon) \end{pmatrix}.
\end{equation}

Substituting these results into equation \eqref{eq_evolution_Hessian_tau}, the explicit form of the second-order sensitivity equations is found to be
\begin{subequations}
	\label{eq:Hessian_system}
	\begin{align}
		&({\Theta}^1_{ab})_{,n} = {\Theta}^2_{ab},\\
		&({\Theta}^2_{ab})_{,n} + g_{,\phi} (3-\epsilon ) {\Theta}^1_{ab} +  \big((3-\epsilon )-\Pi (g+\Pi)\big) {\Theta}^2_{ab} \nonumber \\
		&+ g_{,\phi\phi} (3-\epsilon ) {J}^1_a {J}^1_b - g_{,\phi} \Pi  ({J}^1_a {J}^2_b + {J}^2_a {J}^1_b) - (g+3 \Pi) {J}^2_a {J}^2_b  \nonumber \\
		&= \frac{g k^2 \Pi}{a^2 H^2} ({C}_a {J}^1_b + {C}_b {J}^1_a) + \frac{3 k^2 (\epsilon -1)}{a^2 V} ({C}_a {J}^2_b + {C}_b {J}^2_a).
	\end{align}
\end{subequations}
Therefore, the second-order sensitivity is driven by the first-order Jacobians and the background geometry, effectively mapping the nonlinear dynamics of the field perturbations into a linear system of differential equations. This approach facilitates the derivation of analytic results for $f_{\rm NL}$ in both standard and gradient-corrected $\delta N$ scenarios. The implementation of this formalism in the analytical calculation of non-Gaussianity is discussed in Section \ref{SO_CSA_Staro}.

\subsection{Equivalence of CSA and Linear Perturbation Theory}
\label{Equivalence_CSA_MS}

We now demonstrate that the power spectrum obtained via the CSA method is equivalent to the result obtained from the MS equation. To this end, we multiply the sensitivity equation~\eqref{SO_J_evolution} for $J^1_1$ by $\delta X^1$ and the corresponding equation for $J^1_2$ by $\delta X^2$, and then add the two resulting expressions. This yields
\begin{equation}
	(J^1_a \delta X^a)_{,nn} - \Big[ \Pi(g + \Pi) + (\epsilon - 3) \Big] (J^1_a \delta X^a)_{,n} + (3 - \epsilon )g_{,\phi} (J^1_a \delta X^a) = -\frac{k^2}{a^2 H^2} \Pi C_a \delta X^a .
\end{equation}
Next, we employ the relation
\begin{equation}
	C_a \delta X^a
	= \frac{\partial \hat{\mathcal{R}}}{\partial X^a} \, \delta X^a
	= \hat{\mathcal{R}}  ,
\end{equation}
which, together with \eqref{linear_dN_expansion_tau}, leads to
\begin{equation}
	\delta N_{,nn} + \left(3 - \epsilon + \frac{3 \eta }{3 - \epsilon} \right) \delta N_{,n} - \frac{k^2 }{a^2 H^2} \hat{\mathcal{R}} = 0.
	\label{compare_CSA}
\end{equation}
In the limit $\epsilon \ll 1$, and provided that all gradient contributions are properly included in the evaluation of $\hat{\mathcal{R}}$, equation~\eqref{compare_CSA} indicates that $\delta N$ evolves in the same manner as $-\mathcal{R}$ obtained from the exact MS equation~\eqref{exact_MS_N}. It is also necessary to compare the initial conditions of this equation with those of the MS equation. One can verify that the initial condition matrix $J_*$ reproduces the correct MS initial conditions after applying the change of variables discussed above. Equation \eqref{compare_CSA} also indicates that the $\delta N$ formalism begins to deviate from the MS result in scenarios where $\epsilon$ becomes large. We have verified this behavior numerically in the punctuated inflation model~\cite{jain2008double, jain2010tensor, hazra2013bingo}, which includes a short period of large $\epsilon$. The corresponding results are presented in Appendix~\ref{appen_punctuated_inflation}.

We must clarify the origin of the additional terms in equation~\eqref{compare_CSA}, given that the equivalence between the perturbed background equation and linear perturbation theory demonstrated in Section~\ref{sec_dn_with_gradients} was exact. As established in earlier works~\cite{kodama1998evolution, sasaki1998super}, strictly adhering to the SUA inherently omits the momentum constraint. The momentum constraint, derived from the Einstein equations, contains spatial derivatives on both sides, evaluating to a trivial identity within the strict SUA framework. In deriving the perturbed KG equation~\eqref{perturbed_BG}, this constraint is instead imported from perturbation theory. Physically, the momentum constraint encapsulates information regarding the interactions between distinct FLRW patches; thus, incorporating it constitutes a departure from the exact SUA. Nevertheless, as noted in the original work~\cite{pattison2019stochastic}, employing the spatially flat gauge justifies its use in the limit of small $\epsilon$, where the perturbation of the lapse function, $A \propto \dot{\phi} \delta \phi$, vanishes and the local proper time of each patch synchronizes with the global coordinate time. If $\epsilon$ is non-negligible, the SUA equation~\eqref{perturbed_BG} cannot be mapped to perturbation theory unless local variations in the proper time of the patches are properly accounted for. Consequently, the equivalence between the gradient-corrected $\delta N$ formalism and perturbation theory, as shown in Section~\ref{sec_dn_with_gradients}, holds strictly in the limit $\epsilon \rightarrow 0$.

By working exclusively with linear equations within the SUA and neglecting the momentum constraint, one can determine the exact evolution of the gauge-invariant MS variable, or the comoving curvature perturbation. However, as demonstrated in Reference~\cite{cruces2022review}, this approach introduces residual terms proportional to $\epsilon \eta$ into the evolution equation for $Q$, deviating from the standard linear perturbation theory result~\eqref{perturbed_KG}. This discrepancy illustrates that naively imposing the momentum constraint from perturbation theory onto the SUA equations results in the loss of subtle dynamical information. This reinforces the conclusion that the long-wavelength equivalence between the SUA and linear perturbation theory is robust only in the small-$\epsilon$ regime.

In the gradient-corrected $\delta N$ formalism, a source term is introduced to mimic the gradient interactions absent in the exact SUA. Nevertheless, this framework continues to treat patches as isolated universes, each governed by the corrected KG equation~\eqref{KG_with_source}. Because the momentum constraint remains absent, local quantities cannot be entirely mapped to the global background coordinate system. When the momentum constraint is enforced, all patches become interdependent, constrained by a unified global condition. While the gradient interaction source term, $\mathcal{S}$, successfully captures the spatial interactions of the field perturbations, it is not designed to replicate the global effects of the momentum constraint. This fundamental difference explains why additional terms dependent on the slow-roll parameters emerge when comparing the exact output of the $\delta N$ formalism~\eqref{compare_CSA} with the MS equation \eqref{exact_MS_N}; these terms are precisely the remnants of the omitted momentum constraint.

To demonstrate this point explicitly, Appendix~\ref{sec_dN_source_SUA} derives the evolution equation for $\mathcal{R}$ using an exact linear analysis within the SUA, where the momentum constraint is absent. In this case, additional corrections of order $\epsilon$ naturally arise in the evolution equation. We show that the output of the CSA formulation is fully consistent with these perturbative results, thereby confirming that the small discrepancies between the CSA result and the MS equation originate from the missing momentum constraint.

As a further consistency check of the accuracy and versatility of the CSA framework, Appendix~\ref{sec_dN_sc_SUA} derives the evolution equation for $\mathcal{R}$ within the $\delta N$ formalism that incorporates spatial curvature, as introduced in Ref.~\cite{artigas2025extended}. Using both an exact SUA calculation and the CSA approach, we demonstrate that the two methods yield identical results. This agreement validates the robustness of CSA-based calculations across different implementations of the $\delta N$ formalism.

The influence of the missing momentum constraint can be systematically incorporated into the formalism. Inspecting the final result~\eqref{compare_CSA}, it becomes evident that the exact equivalence between the CSA and the linear perturbation theory is restored by a redefinition of the source term in the form
\begin{equation}
	\mathcal{S} \rightarrow \mathcal{S} + \tilde{\mathcal{S}},
	\label{corrected_source_term}
\end{equation}
with
\begin{equation}
	\tilde{\mathcal{S}} = \tilde{B} \hat{\mathcal{R}}_{,n}, \qquad \tilde{B} \equiv - \Pi^2 (g + \Pi).
\end{equation}
This source term modifies the first-order forcing tensor as
\begin{equation}
	\left[\Sigma^i_a\right] \rightarrow \left[\Sigma^i_a\right] + 
	\tilde{B} \begin{pmatrix} 0 & 0 \\ ({C}_1)_{,n} & ({C}_2)_{,n} \end{pmatrix},
\end{equation}
and the second-order forcing as
\begin{align}
	\Sigma^2_{ab} \rightarrow \Sigma^2_{ab} +  (\tilde{B}_j J^j_a) ({C}_b)_{,n} + ( \tilde{B}_j J^j_b ) ({C}_a)_{,n},
\end{align}
where $\tilde{B}_j \equiv \partial \tilde{B}/\partial Y^j$ is given by:
\begin{equation}
	\tilde{B}_j = - \Pi \begin{pmatrix} \Pi g_{,\phi} \\ 2 g + 3 \Pi \end{pmatrix}.
\end{equation}

In Appendix~\ref{appen_punctuated_inflation}, where we analyze the breakdown of the $\delta N$ formalism during punctuated inflation, we also evaluate and plot the power spectrum obtained using this corrected source term. As illustrated in Figure~\ref{Punctuated_inflation}, the modified source term~\eqref{corrected_source_term} accurately reproduces the predictions of linear perturbation theory even during phases with transiently large $\epsilon$, thereby validating the proposed correction. Although the impact of the correction terms in~\eqref{corrected_source_term} is highly suppressed in scenarios with small $\epsilon$, we still retain the corrected source term in our numerical evaluations of the Starobinsky model in Section \ref{sec_example_staro} to capture any subtle variations in the second-order results. A sample source code demonstrating these evaluations is publicly available on GitHub \cite{github_code}. For our analytical calculations, we adopt the small-$\epsilon$ approximation, and these corrections are consequently neglected.

Let us briefly comment on the numerical implementation of the CSA method. In standard implementations of the $\delta N$ formalism, as expressed in equation \eqref{dN_expansion_standard}, one must evaluate the derivatives of the number of $e$-folds realized between an initial spatially flat hypersurface and a final uniform-density hypersurface. In numerical computations, evaluating these derivatives necessitates solving the background evolution between the two hypersurfaces using at least two distinct initial conditions. For the full-gradient $\delta N$ formalism discussed in Section~\ref{review_dN_fullgrad}, the background field equation must be solved alongside a source term constructed from $\hat{\mathcal{R}}$, which is obtained by solving the MS equation or gradient expansion. As demonstrated in the public source code provided by Ref.~\cite{ahmadi2026delta}, while such evaluations can be executed relatively quickly, they frequently suffer from a degradation in numerical accuracy.

The CSA formulation significantly streamlines these numerical evaluations through several key advantages. First, rather than repeatedly solving the background evolution, CSA reduces the problem to determining the continuous sensitivity of the field values with respect to their initial conditions. This approach requires solving a system of coupled first-order differential equations, which is highly computationally stable and efficient. Second, by employing a simple identification that we will later introduce in equation~\eqref{identification_full}, we can entirely circumvent the numerical difficulties associated with constructing the source term from either the gradient expansion or the MS equation. Finally, because the $\delta N$ formalism with full gradient interactions completely captures the non-adiabatic evolution of curvature perturbations, we can match the Jacobian directly to the Bunch-Davies vacuum deep inside the horizon. This effectively eliminates the need to integrate the MS equation up to a matching surface in order to match the $\delta N$ framework with the MS solution.

\section{Application to the Starobinsky Model}
\label{sec_example_staro}

To illustrate the practical utility of the CSA formalism developed in the previous section, we now apply it to the Starobinsky model \cite{starobinskij1992spectrum, martin2012scalar, martin2014sharp, ahmadi2022quantum}. This model serves as a paradigmatic example of inflationary dynamics featuring a sharp transition to a non-slow-roll phase, making it an ideal testing ground for formalisms that track the continuous evolution of perturbations. Our primary aim in this section is to evaluate the background dynamics through this sharp feature and to utilize the CSA framework to analytically determine the power spectrum and non-Gaussianity parameter.

\subsection{Background Dynamics and Linear Perturbations}
The Starobinsky model is characterized by a piecewise linear potential with a sudden change in its slope at a specific field value $\phi_T$. The potential can be well-approximated as
\begin{equation}
	V(\phi) =
	\begin{cases}
		V_0 + A_{+}(\phi - \phi_T), & \phi \geq \phi_T, \\[6pt]
		V_0 + A_{-}(\phi - \phi_T), & \phi < \phi_T,
	\end{cases}
	\label{eq_starobinsky_potential}
\end{equation}
where the subscripts ``$+$'' and ``$-$'' denote the pre-transition ($\tau \le \tau_T$) and post-transition ($\tau > \tau_T$) epochs, respectively. We assume that the vacuum energy strictly dominates the potential, $V_0 \gg A_\pm |\phi - \phi_T|$, which places the background in a de Sitter limit where the Hubble parameter $H$ is effectively constant. 

Under the constant $H$ approximation, the KG equation can be integrated analytically. By matching the scalar field $\phi$ and its conformal time derivative $\phi'$ at the comoving transition time $\tau_T$, one obtains the full background evolution. Defining the comoving scale that exits the horizon exactly at the transition as $k_T = -1/\tau_T$, and letting $\Delta A \equiv A_- - A_+ < 0$ represent the discontinuity in the potential's gradient, we can analytically derive the behavior of the Hubble flow parameters.

The first Hubble flow parameter evaluates to
\begin{equation}
	\epsilon(\tau) = 
	\begin{cases}
		\dfrac{A_+^2}{18 H^4}, & \text{for } \tau \le \tau_T, \\[6pt]
		\dfrac{(\Delta A k_T^3 \tau^3 + A_-)^2}{18 H^4}, & \text{for } \tau > \tau_T.
	\end{cases}
	\label{epsilon1_starobinsky}
\end{equation}
Because the field velocity $\phi'$ is continuous across the transition, $\epsilon$ remains perfectly continuous at $\tau_T$. However, the second flow parameter, $\eta$, relies on the second derivative of the field. This parameter is evaluates to
\begin{equation}
	\eta(\tau) = 
	\begin{cases}
		0, & \text{for } \tau \le \tau_T, \\[6pt]
		-\dfrac{6 \Delta A k_T^3 \tau^3}{\Delta A k_T^3 \tau^3 + A_-}, & \text{for } \tau > \tau_T.
	\end{cases}
	\label{epsilon2_starobinsky}
\end{equation}
The jump in the potential's derivative induces a sudden discontinuity in $\eta$. This parameter abruptly transitions from zero to a finite negative value, signifying a temporary, yet severe, breakdown of the standard slow-roll approximation. This period, during which $\eta$ drops to a value of approximately $-6$, corresponds to the well-known ultra-slow-roll (USR) phase \cite{tsamis2004improved, kinney2005horizon, namjoo2013violation, martin2013ultra}. Both the slow-roll and USR phases are recognized as special cases of a broader class of models known as constant-roll inflation, and the transition between these two regimes is referred to as constant-roll duality \cite{martin2013ultra, motohashi2015inflation, odintsov2017inflation, anguelova2018systematics, morse2018large, yi2018constant, lin2019dynamical, ghersi2019observational, mohammadi2023constant, mohammad2024analytical}.

Using the field dynamics dictated by the background equations, the evolution of $z = \phi'/H$ as a function of conformal time is given by
\begin{equation}
	z(\tau) = \text{sign}(\phi') 
	\begin{cases}
		\dfrac{A_+}{3 H^3} \tau^{-1}, & \text{for } \tau \le \tau_T, \\[6pt]
		\dfrac{A_- + \Delta A k_T^3 \tau^3}{3 H^3} \tau^{-1}, & \text{for } \tau > \tau_T.
	\end{cases}
	\label{z_starobinsky}
\end{equation}
Inspection of this relation reveals that $z$ is continuous at the transition. However, its conformal time derivative, $z'$, inherits a step discontinuity proportional to the jump in the potential's slope.

To find the evolution of the curvature perturbation $\mathcal{R}$ before and after the transition, it is convenient to define the MS variable $v \equiv a Q = z \mathcal{R}$. This allows us to write the MS equation \eqref{MS_equation_R} in the canonical form
\begin{equation}
	v'' + \left(k^2 - \frac{z''}{z}\right) v=0,
	\label{MS_nu_equation}
\end{equation}
where the effective potential $z''/z$ is generally expressed in terms of the Hubble flow parameters as
\begin{equation}
	\frac{z^{\prime\prime}}{z} = a^2 H^2 \left(2- \epsilon +\frac{3 }{2} \eta +\frac{1}{4} \eta^2-\frac{1}{2}\epsilon \eta+ \frac{1}{2} \eta \, \xi \right).
\end{equation}
Assuming the first Hubble flow parameter is small ($\epsilon \ll 1$), and using the relation $\eta \, \xi \approx -3 \eta - \eta^2 / 2$ characteristic of the linear potential, the effective potential can be approximated as $z''/z \approx 2 / \tau^2$ for both the USR and slow-roll phases. Consequently, the mode solutions for both stages of evolution can be expressed as a linear combination of de Sitter modes:
\begin{equation}
	v_{\pm} (\tau) = \frac{\mathcal{B}_{1\pm}}{\sqrt{2k}} \left( 1 - \frac{i}{k \tau} \right) e^{-i k \tau} + \frac{\mathcal{B}_{2\pm}}{\sqrt{2k}} \left( 1 + \frac{i}{k \tau} \right) e^{i k \tau}.
	\label{pre_transition_nu}
\end{equation}
In the pre-transition phase, we naturally assume the modes originate in the Bunch-Davies vacuum, which fixes the initial Bogoliubov coefficients to
\begin{equation}
	\mathcal{B}_{1+} = 1, \quad \mathcal{B}_{2+} = 0.
	\label{B_BD_coifficients}
\end{equation}
After the transition, the Bogoliubov coefficients must be evaluated from the boundary conditions at $\tau_T$. Using the relations \eqref{z_starobinsky}, the effective potential across the transition is derived as
\begin{equation}
	\frac{z''}{z} = \frac{2}{\tau^2} + \frac{3 \Delta A k_T}{A_+} \delta (\tau - \tau_T).
	\label{zppoz_jump}
\end{equation}
Integrating the MS equation \eqref{MS_nu_equation} across an infinitesimal interval $[\tau_T - \varepsilon, \tau_T + \varepsilon]$, we obtain the boundary condition governing the discrete jump in the derivative of the mode function:
\begin{equation}
	\Delta(v') \equiv v'(\tau_T^+) - v'(\tau_T^-) = \frac{3 \Delta A k_T}{A_+} v(\tau_T).
\end{equation}
By imposing the continuity of $v$ and the derived jump condition for $v'$, the Bogoliubov coefficients for the post-transition epoch are found to be
\begin{equation}
	\mathcal{B}_{1-} = 1 + \frac{3i}{2} \frac{\Delta A k_T}{A_+ k} \left( 1 + \frac{k_T^2}{k^2} \right), \qquad
	\mathcal{B}_{2-} = -\frac{3i}{2} \frac{\Delta A k_T}{A_+ k} \left( 1 + \frac{ik_T}{k} \right)^2 e^{2ik/k_T}. \label{beta_post}
\end{equation}
Consequently, the curvature perturbation at late times can be evaluated as
\begin{equation}
	\mathcal{R}(k) = \frac{3 H^3}{2 \sqrt{2} A_- A_+ k^{9/2}} \left[ -2 i A_+ k^3 + 3 k_T \Delta A \left( k^2 + k_T^2 + e^{\frac{2 i k}{k_T}} (k + i k_T)^2 \right) \right],
	\label{R_star}
\end{equation}
yielding the dimensionless power spectrum
\begin{align}
	\mathcal{P}_{\mathcal{R}}(k) &\equiv \frac{k^3}{2 \pi^2} \left| \mathcal{R}(k) \right|^2 \nonumber \\
	&= \frac{9 H^6}{8 \pi^2 A_-^2 A_+^2 \tilde{k}^6} \Bigg[ 2 A_+^2 \tilde{k}^6 + 9 \Delta A^2 (\tilde{k}^2 + 1)^2 \nonumber \\
	&\quad + 3 \Delta A \Big( 3 \Delta A (\tilde{k}^4 - 1) - 4 A_+ \tilde{k}^4 \Big) \cos(2\tilde{k}) \nonumber \\
	&\quad - 6 \Delta A \tilde{k} \Big( A_+ \tilde{k}^2 (\tilde{k}^2 - 1) + 3 \Delta A (\tilde{k}^2 + 1) \Big) \sin(2\tilde{k}) \Bigg],
	\label{Power_MS}
\end{align}
where $\tilde{k} \equiv k / k_T$ is defined for simplicity. The MS solution derived here will enter the CSA of the $\delta N$ formalism through the matching procedure.

\subsection{First-Order Sensitivities}
\label{FO_CSA_Staro}

In this section, we derive the first-order sensitivities for the Starobinsky model. To systematically construct the full solution, we first evaluate the homogeneous part of the linear sensitivity equations. This allows us to recover the results of the standard $\delta N$ formalism. Subsequently, these homogeneous solutions serve as the foundation for constructing the full Green's function, which incorporates the effects of gradient interactions. In particular, we evaluate the leading $\mathcal{O}(k^{2})$ spatial gradient corrections using the Green's function method. Finally, we incorporate the full spatial gradient interactions into the sensitivity equations, proving that this framework can analytically recover the exact MS solution independently of the chosen matching time. For mathematical convenience throughout this derivation, we switch to conformal time.

\subsubsection{Homogeneous Jacobian}
\label{dn_homogeneous}

We now analytically solve the homogeneous part of the linear sensitivity equations \eqref{SO_J_evolution} to recover the results of the standard $\delta N$ formalism. These homogeneous solutions can subsequently be used within a Green's function integral to incorporate gradient interactions, as we demonstrate in the next section.

In this section, we will find it easier to work with conformal time as the time variable. In the $\epsilon \to 0$ limit relevant to the Starobinsky background, the first-order sensitivity equations \eqref{SO_J_evolution} reduce to
\begin{equation}
	(\bar{J}^1_a)''- \frac{2}{\tau} (\bar{J}^1_a)' + a^2 V_{,\phi\phi} \bar{J}^1_a = 0,
	\label{J_evolve_star}
\end{equation}
where we use an overbar to indicate homogeneous solutions. During both the pre- and post-transition stages of the Starobinsky model, the second derivative of the potential, $V_{,\phi\phi}$, vanishes exactly. Consequently, the general homogeneous solution takes the form
\begin{equation}
	\bar{J}^1_{a\pm}(\tau) = \alpha_{a\pm} \tau^3 + \beta_{a\pm}.
	\label{HMJ_solution_staro}
\end{equation}
To determine the boundary conditions for the derivatives of $\bar{J}^1_a$, we must account for the discontinuity in the potential's slope at the transition, which induces a Dirac delta function in the second derivative:
\begin{equation}
	V_{,\phi\phi} = -\frac{\Delta A}{|\phi'_T|} \delta(\tau - \tau_T).
\end{equation}
Substituting this expression into equation \eqref{J_evolve_star} and integrating across an infinitesimal conformal time interval $[\tau_T - \varepsilon, \tau_T + \varepsilon]$ yields the matching condition for the derivative jump:
\begin{equation}
	\Delta (\bar{J}^1_{a})' = \frac{a_T^2 \Delta A}{|\phi'_T|} \bar{J}^1_{a}(\tau_T).
	\label{boundary_J_star}
\end{equation}

For modes that cross the transition after the matching time, $k \le \sigma k_T$, the background evolves on the $A_+$ branch until the junction. We adopt $X=(\mathcal{R}_*, \mathcal{R}^{\prime}_*)$ as the set of independent initial conditions. Consequently, the initial conditions for the Jacobian matrix elements are provided by equation~\eqref{Js_tau}. Applying these, the constants of integration for the pre-transition stage are uniquely determined to be:
\begin{subequations}\label{HM_ab_1}
\begin{align}
	\alpha_{1+} &= 0, & \beta_{1+} &= -\frac{A_+}{3 H^2}, \\
	\alpha_{2+} &= -\frac{A_+ k^2}{9 H^2 \sigma^2}, & \beta_{2+} &= -\frac{A_+ \sigma}{9 H^2 k},
\end{align}
\end{subequations}
where we have utilized \eqref{z_starobinsky} to substitute the field velocity $\phi' = z H$. By applying the boundary conditions across the transition, the integration constants for the second stage are evaluated as
\begin{subequations}\label{HM_ab_2}
	\begin{align}
		\alpha_{1-} &= -\frac{\Delta A k_T^3}{3 H^2}, & \beta_{1-} &= -\frac{A_-}{3 H^2},  \\
		\alpha_{2-} &=-\frac{A_- k^2}{9 H^2 \sigma ^2}+\frac{2 \Delta A k^2}{9 H^2 \sigma ^2}-\frac{\Delta A k_T^3 \sigma }{9 H^2 k}, & \beta_{2-} &=  \frac{\Delta A k^2}{9 H^2 k_T^3 \sigma^2} - \frac{A_- \sigma}{9 H^2 k}.
	\end{align}
\end{subequations}

At late times, as conformal time approaches zero, the Jacobian elements in equation~\eqref{HMJ_solution_staro} asymptotically approach their respective constant terms, $\bar{J}^1_{a-}(\tau_e) \approx \beta_{a-}$. The first-order $\delta N$ expansion is then given by
\begin{align}
	\delta N &= -\frac{1}{\Pi_e}\Bigl[\bar{J}^1_{1}(\tau_e)\,\mathcal{R}_{*} + \bar{J}^1_{2}(\tau_e)\,\mathcal{R}'_{*}\Bigr] \nonumber \\
	&= -\mathcal{R}_{*} - \left[1 - \frac{\Delta A}{A_-}\left(\frac{\tilde{k}}{\sigma}\right)^3\right] \frac{\dot{\mathcal{R}}_{*}}{3H}.
	\label{Standard_dN_1}
\end{align}
For modes that reach the transition before the matching time, $k > \sigma k_T$, the background evolution occurs entirely on the post-transition $A_-$ branch. By matching the homogeneous solution $\bar{J}^1_{a-}$ directly to the initial conditions $J_*$ at the matching surface, we find
\begin{subequations}\label{HM_ab_3}
	\begin{align}
		\alpha_{1-} &= -\frac{\Delta A k_T^3}{3 H^2}, & \beta_{1-} &= -\frac{A_-}{3 H^2},  \\
		\alpha_{2-} &= \frac{\Delta A k_T^3 \sigma }{9 H^2 k}-\frac{A_- k^2}{9 H^2 \sigma ^2}, & \beta_{2-} &=  \frac{\Delta A k_T^3 \sigma^4}{9 H^2 k^4} - \frac{A_- \sigma}{9 H^2 k},
	\end{align}
\end{subequations}
which consequently yields the perturbation
\begin{equation}
	\delta N = -\mathcal{R}_{*} - \left[1 - \frac{\Delta A}{A_-}\left(\frac{\sigma}{\tilde{k}}\right)^3\right] \frac{\dot{\mathcal{R}}_{*}}{3H}.
	\label{Standard_dN_2}
\end{equation}

\begin{figure}
	\centering
	\includegraphics[width=0.65\textwidth]{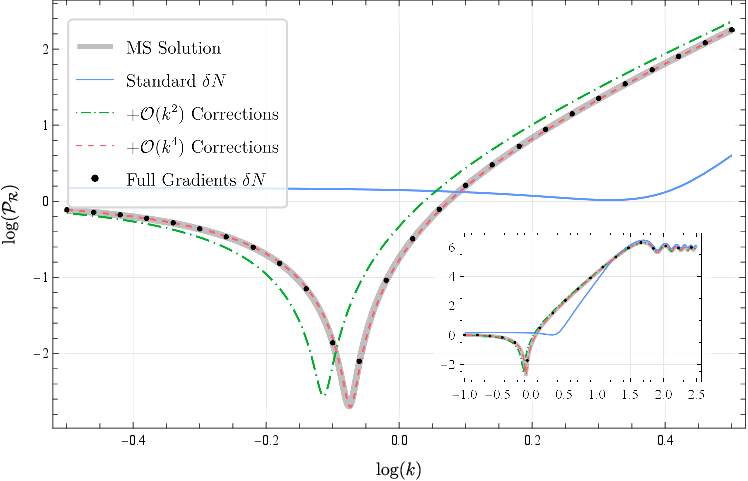}
	\caption{The power spectrum of the Starobinsky model evaluated using the analytic solution of the MS solution \eqref{Power_MS}, standard $\delta N$ formalism results, \eqref{Standard_dN_1} and \eqref{Standard_dN_2}, $\delta N$ with leading gradient corrections \eqref{GC_J_1} and \eqref{GC_J_2}, with inclusion of $\mathcal{O}(k^4)$ corrections \eqref{k4_corrections} and \eqref{k4_corrections2}, and with full gradient interactions \eqref{J1a_full_gradients} and \eqref{J1a_full_gradients2}, as a function of wavenumber. The matching time is chosen at the horizon crossing time, $\sigma = 1$. The numerical values of the model parameters are chosen as $V_0 = 0.137$, $A_+ = 4.56 \times 10^{-3}$, $A_- = 1.139 \times 10^{-3} A_+$, $\phi_T = 0$, and $\phi_{\rm in} = 0.4$. The exact MS solution is denoted by the thick gray line. The standard $\delta N$ approximation, its $\mathcal{O}(k^2)$ corrections, and $\mathcal{O}(k^4)$ corrections are represented by the solid blue, dot-dashed green, and dashed red lines, respectively. The full gradient $\delta N$ result is shown using black bullets. The inset panel illustrates these same quantities over an extended range of wavenumbers.}
	\label{powespectrum_starobinsky_model}
\end{figure}

Together, equations \eqref{Standard_dN_1} and \eqref{Standard_dN_2} fully determine the power spectrum within the standard $\delta N$ approximation. These findings are in perfect agreement with the results of Ref.~\cite{jackson2024separate}, where the authors evaluated the spectrum using the homogeneous matching method and the traditional formulation of the $\delta N$ formalism. The resulting power spectrum is illustrated in Figure~\ref{powespectrum_starobinsky_model}. While it is already apparent that the CSA approach streamlines the standard $\delta N$ calculation, deriving the homogeneous solutions alone does not fully demonstrate its power. In the following section, we proceed to compute the gradient corrections to these homogeneous solutions, thereby showcasing the distinct analytical advantages of the CSA framework.

\subsubsection{Leading Gradient Corrections to Jacobian}
\label{k2_jacobian}

Examining the first-order sensitivity equation~\eqref{SO_J_evolution}, we see that the gradient corrections appear as a source term in a second-order linear ordinary differential equation. This observation naturally suggests that the gradient corrections to the homogeneous solutions derived in the previous section can be evaluated using the Green's function method. The Green's function representation explicitly separates the homogeneous evolution from the gradient-sourced contribution. As a result, it provides a convenient framework for analyzing how the forcing terms modify the Jacobian evolution. In this section, we apply this method to compute the leading gradient corrections, including both adiabatic and non-adiabatic contributions.

In the small $\epsilon$ limit the sensitivity equations \eqref{SO_J_evolution} can be written in the form
\begin{equation}
	(J^1_a)''
	+ p(\tau) (J^1_a)'
	+ q(\tau) J^1_a
	= \tilde{\Sigma}^2_a(\tau),
	\label{green_ODE}
\end{equation}
where
\begin{align}
	p(\tau) = 2 a H, 
	\qquad
	q(\tau) = a^2 V_{,\phi\phi},
	\qquad
	\tilde{\Sigma}^2_a = a^2 H^2 {\Sigma}^2_a = -k^2 \Pi C_a.
\end{align}
Let \(\bar{J}^1_1\) and \(\bar{J}^1_2\) denote the homogeneous solutions corresponding to \(J^1_1\) and \(J^1_2\). Since the homogeneous parts of the sensitivity equations for \(J^1_a\) are identical, the solutions \(\bar{J}^1_1\) and \(\bar{J}^1_2\) form a complete set of independent solutions relevant for the Green's function method. Therefore, the gradient-corrected solutions for the Jacobian components can be written as
\begin{equation}
	J^i_a = \bar{J}^i_a + \bar{J}^i_k I^k_a,
	\label{general_Green_function_solution}
\end{equation}
where
\begin{equation}
	I^k_a \equiv \int_{\tau_*}^{\tau}\!  \frac{ \varepsilon^{k l}  \bar{J}^1_{l} \tilde{\Sigma}^2_a}{W} \,d\tilde{\tau}.
\end{equation}
Here $\varepsilon^{kl}$ is the Levi--Civita symbol, and
\begin{equation}
	W = \bar{J}_1 \bar{J}'_2 - \bar{J}_2 \bar{J}'_1,
\end{equation}
is the Wronskian associated with the two homogeneous solutions. The Wronskian satisfies the identity $W' = -p(\tau) W$. Integrating this relation yields the useful relation
\begin{equation}
	W(\tau) =
	W(\tau_*)
	\exp\!\left[
	-
	\int_{\tau_*}^{\tau}\!
	p(\tilde{\tau})\, d\tilde{\tau}
	\right].
	\label{Wronskian_main_formula}
\end{equation}

Let us first consider modes that reach the matching surface before the transition ($k < \sigma k_T$). As previously established, since the homogeneous differential equations governing the Jacobian components $J^1_a$ are identical, we can utilize the homogeneous solutions \eqref{HMJ_solution_staro} to construct the gradient-corrected solution. Following \eqref{general_Green_function_solution}, the late-time gradient-corrected Jacobian takes the form:
\begin{equation}
	J^1_{a} (\tau_e)
	=
	\bar{J}^1_{a} (\tau_e)
	-
	\bar{J}^1_{1} (\tau_e)
	\int_{\tau_*}^{\tau_e}
	\frac{\bar{J}^1_{2} \tilde{\Sigma}^{2}_a}{W}
	\, d\tilde{\tau}
	+
	\bar{J}^1_{2} (\tau_e)
	\int_{\tau_*}^{\tau_e}
	\frac{\bar{J}^1_{1} \tilde{\Sigma}^{2}_a}{W}
	\, d\tilde{\tau}.
	\label{Green_function_Jij}
\end{equation}
Splitting the integral into pre- and post-transition epochs and inserting the homogeneous solutions \eqref{HMJ_solution_staro}, we obtain:
\begin{equation}
	J^1_a (\tau_e)
	= \beta_{a-}
	+ \mathcal{C}_1 \int_{\tau_*}^{\tau_T}
	\frac{\tilde{\Sigma}^{2}_{a+} \tilde{\tau}^3}{W} 
	\, d\tilde{\tau}
	+ \mathcal{C}_2 \int_{\tau_*}^{\tau_T}
	\frac{\tilde{\Sigma}^{2}_{a+}}{W}
	\, d\tilde{\tau}
	+ \mathcal{C}_3 \int_{\tau_T}^{\tau_e}
	\frac{\tilde{\Sigma}^{2}_{a-} \tilde{\tau}^3}{W} 
	\, d\tilde{\tau},
	\label{Green_func_k2_correction}
\end{equation}
where the combined prefactors $\mathcal{C}_i$ are defined as:
\begin{subequations}\label{Ci_constants}
	\begin{align}
		\mathcal{C}_1 &\equiv \alpha_{1+} \beta_{2-} - \alpha_{2+} \beta_{1-} = -\frac{A_- A_+ k^2}{27 H^4 \sigma^2}, \\
		\mathcal{C}_2 &\equiv \beta_{1+} \beta_{2-} - \beta_{2+} \beta_{1-}  = -\frac{A_+ \Delta A k^2}{27 H^4 k_T^3 \sigma^2}, \\
		\mathcal{C}_3 &\equiv \alpha_{1-} \beta_{2-} - \alpha_{2-} \beta_{1-} = -\frac{A_+^2 k^2}{27 H^4 \sigma^2}.
	\end{align} 
\end{subequations}
Here, equations \eqref{HM_ab_1} and \eqref{HM_ab_2} dictate the integration constants for the $k < \sigma k_T$ regime. The Wronskian remains continuous across the transition and is evaluated using relation \eqref{Wronskian_main_formula} as:
\begin{equation}
	W(\tau) =
	W_*
	\exp\!\left[
	\int_{\tau_*}^{\tau}
	\frac{2}{\tilde{\tau} }\, d\tilde{\tau}
	\right] = W_* \frac{\tau^2}{\tau_*^2},
	\label{wronskian_green}
\end{equation}
with the initial value defined by:
\begin{equation}
	W_* = \left. \left[\bar{J}^1_{1+} (\bar{J}^1_{2+})' - \bar{J}^1_{2+} (\bar{J}^1_{1+})'\right] \right|_{\tau_*} = \frac{A_+^2}{9 H^4} \ .
\end{equation}

The forcing term $\tilde{\Sigma}^{2}_a$, however, differs between the two stages of the model. From relation~\eqref{homoheneous_hatCD}, we note that the mode functions at leading order take the form:
\begin{equation}
	u_{\rm ad} (\tau) = 1, \qquad u_{\rm nad} (\tau) = z_*^2
	\int_{\tau_*}^{\tau}\!
	\frac{d\tilde{\tau}}{z^2(\tilde{\tau})}.
\end{equation}
This implies $C_{1\pm} = 1$, while for the non-adiabatic component we find:
\begin{subequations}
	\begin{align}
		C_{2+} &= z_{+*}^2
		\int_{\tau_*}^{\tau}\!
		\frac{d\tilde{\tau}}{z_{+}^2} = \frac{k^3 \tau ^3+\sigma ^3}{3 k \sigma ^2}, \\
		C_{2-} &= z_{+*}^2 \left[ \int_{\tau_*}^{\tau_T}\!\!
		\frac{d\tilde{\tau}}{z_{+}^2}  + \int_{\tau_T}^{\tau}\! \frac{d\tilde{\tau}}{z_{-}^2}  \right] \nonumber \\
		&= \frac{\sigma }{3 k} + \frac{k^2}{3 k_T^3 \sigma ^2} \left(\frac{A_+}{\Delta A} -\frac{A_+^2}{\Delta A \left( A_-+\Delta A k_T^3 \tau ^3\right)}-1 \right).
	\end{align}
\end{subequations}
Finally, noting that the forcing components $\tilde{\Sigma}^2_{a\pm}$ can be well-approximated as $\tilde{\Sigma}^2_{a\pm} \approx H k^2 \tau z_{\pm} C_{a\pm}$, we can evaluate the integrals in \eqref{Green_func_k2_correction}:
\begin{subequations}
	\begin{align}
		\int_{\tau_*}^{\tau_T}
		\frac{\tilde{\Sigma}^{2}_{1+} \tilde{\tau}^3}{W} 
		\, d\tilde{\tau} &= \frac{3 H^2 \sigma^2}{2 A_+ k^2} (\tilde{k}^2 - \sigma^2),
		\\
		\int_{\tau_*}^{\tau_T}
		\frac{\tilde{\Sigma}^{2}_{1+}}{W}
		\, d\tilde{\tau} &=  \frac{3 H^2 k_T \sigma}{A_+} (  \sigma - \tilde{k} ),
		\\
		\int_{\tau_T}^{\tau_e}
		\frac{\tilde{\Sigma}^{2}_{1-} \tilde{\tau}^3}{W} 
		\, d\tilde{\tau} &= - \frac{3 H^2 \sigma^2}{10 A_+^2 k_T^2} \left(3 \Delta A + 5 A_+\right),
		\\
		\int_{\tau_*}^{\tau_T}
		\frac{\tilde{\Sigma}^{2}_{2+} \tilde{\tau}^3}{W} 
		\, d\tilde{\tau} &= \frac{H^2}{10 A_+ k^3} ( 5\sigma^3 \tilde{k}^2 - 2\tilde{k}^5 - 3 \sigma^5 ),
		\\
		\int_{\tau_*}^{\tau_T}
		\frac{\tilde{\Sigma}^{2}_{2+}}{W}
		\, d\tilde{\tau} &= \frac{H^2}{2 A_+ \tilde{k}}  (\tilde{k} + 2 \sigma ) (\tilde{k} - \sigma )^2,
		\\
		\int_{\tau_T}^{\tau_e}
		\frac{\tilde{\Sigma}^{2}_{2-} \tilde{\tau}^3}{W} 
		\, d\tilde{\tau} &= \frac{H^2}{10 A_+^2 \tilde{k} k_T^3} \left((2 A_++3 \Delta A) (\tilde{k}^3-\sigma ^3)-3 A_+ \sigma ^3\right) .
	\end{align}
\end{subequations}
Substituting all these evaluations back into \eqref{Green_func_k2_correction}, we find the following expressions for the Jacobians:
\begin{subequations}\label{GC_J_1}
\begin{align}
	\frac{J^1_1(\tau_e)}{\Pi_e}  &= 1 - \frac{\Delta A \tilde{k}^3}{3 A_- \sigma} + \frac{2 \Delta A \tilde{k}^2}{5 A_- } - \frac{\sigma^2}{6}, \\
	\frac{J^1_2(\tau_e)}{\Pi_e}  &= \frac{\sigma}{3 k} \left[1 - \frac{\Delta A \tilde{k}^3}{A_-  \sigma^3} + \frac{\Delta A \tilde{k}^5}{5 A_- \sigma^3}  - \frac{\Delta A \tilde{k}^3}{2 A_- \sigma} + \frac{2 \Delta A \tilde{k}^2}{5 A_- } - \frac{\sigma^2}{10}\right].
\end{align}
\end{subequations}
These expressions can be directly inserted into relation \eqref{linear_dN_expansion_tau} to construct the linear gradient-corrected $\delta N$.

For modes where the matching surface is crossed after the transition ($k > \sigma k_T$), relation \eqref{Green_function_Jij} reduces to:
\begin{equation}
	J^1_a (\tau_e)
	= \beta_{1-}
	+ \mathcal{C}_4 \int_{\tau_*}^{\tau_e}
	\frac{\tilde{\Sigma}^{2}_{a-} \tilde{\tau}^3}{W} 
	\, d\tilde{\tau},
	\label{Green_func_k2_correction_2}
\end{equation}
where the prefactor is given by:
\begin{equation}
	\mathcal{C}_4 \equiv \alpha_{1-} \beta_{2-} - \alpha_{2-} \beta_{1-} = -\frac{k_T^2(A_- \tilde{k}^3 - \Delta A  \sigma^3)^2}{27 H^4 \tilde{k}^4 \sigma^2}.
\end{equation} 
Here, the integration constants from \eqref{HM_ab_3} are applied. The Wronskian retains the same functional form as \eqref{wronskian_green}, but with the initial value adjusted to:
\begin{equation}
	W_* = \left. \left[\bar{J}^1_{1-} (\bar{J}^1_{2-})' - \bar{J}^1_{2-} (\bar{J}^1_{1-})'\right] \right|_{\tau_*} = \frac{(A_- \tilde{k}^3 - \Delta A  \sigma ^3)^2}{9 H^4 \tilde{k}^6}.
\end{equation}
As before, $C_{1-} = 1$, while for $C_{2-}$ we obtain:
\begin{equation}
	C_{2-} = z_{-*}^2 \int_{\tau_*}^{\tau}
	\frac{d\tilde{\tau}}{z_{-}^2} = \frac{\left(k^3 \tau^3 + \sigma^3\right) \left(A_- k^3 - \Delta A k_T^3 \sigma^3\right)}{3 k^4 \sigma^2 \left(A_- + \Delta A k_T^3 \tau^3\right)}.
\end{equation}
Evaluating the integrals in \eqref{Green_func_k2_correction_2} yields:
\begin{subequations}
	\begin{align}
		\int_{\tau_*}^{\tau_e}
		\frac{\tilde{\Sigma}^{2}_{1-} \tilde{\tau}^3}{W} 
		\, d\tilde{\tau} &=  \frac{3 H^2 \tilde{k} \sigma^4(2 \Delta A  \sigma^3 - 5 A_- \tilde{k}^3)}{ 10 k_T^2 (A_- \tilde{k}^3 - \Delta A  \sigma^3)^2}, \\
		\int_{\tau_*}^{\tau_e}
		\frac{\tilde{\Sigma}^{2}_{2-} \tilde{\tau}^3}{W} 
		\, d\tilde{\tau} &= \frac{3 H^2 \sigma^5}{10 k_T^3 (\Delta A \sigma^3 - A_- \tilde{k}^3)}.
	\end{align}
\end{subequations}
Substituting these integrated forms back into the relation \eqref{Green_func_k2_correction_2}, we extract the post-transition Jacobians:
\begin{subequations}\label{GC_J_2}
	\begin{align}
		\frac{J^1_1(\tau_e)}{\Pi_e}  &= 1 +  \frac{\Delta A  \sigma^5}{15 A_- \tilde{k}^3} - \frac{\sigma^2}{6}, \\
		\frac{J^1_2(\tau_e)}{\Pi_e}  &= \frac{\sigma}{3 k} \left[1  - \frac{\Delta A \sigma^3}{A_- \tilde{k}^3} + \frac{\Delta A  \sigma^5}{10 A_- \tilde{k}^3} - \frac{\sigma^2}{10}\right],
	\end{align}
\end{subequations}
which can then be used in relation \eqref{linear_dN_expansion_tau} to construct the linear $\delta N$ results for the $k > \sigma k_T$ regime. The resulting power spectrum, calculated using the gradient-corrected relations \eqref{GC_J_1} and \eqref{GC_J_2}, is depicted in Figure \ref{powespectrum_starobinsky_model}, where it is compared with the standard $\delta N$ results and the analytic MS solution.

The gradient-corrected $\delta N$ formalism introduces feature-independent corrections (specifically the $-\sigma^2/6$ and $-\sigma^2/10$ terms) that physically capture the residual evolution of modes immediately following horizon crossing. Standard $\delta N$ relies on the SUA, assuming that spatial gradients completely vanish and curvature perturbations freeze instantly at $k = aH$. However, the exact Mukhanov-Sasaki equation dictates a continuous, rather than instantaneous, relaxation phase. The pure $\sigma^2$ terms act as an intrinsic sub-horizon memory effect, correcting the artificial "sudden freeze" assumption by accounting for the residual damped oscillation of the mode. Because these terms appear even in the absence of a sharp transition ($\Delta A = 0$), they represent a fundamental baseline correction to the separate universe limit.

Beyond this baseline relaxation, the formalism yields transition-induced corrections that explicitly couple the sudden background discontinuity ($\Delta A$) to the residual spatial curvature of the perturbation. The physical manifestation of this coupling depends on the scale of the mode relative to the transition. For modes reaching the matching surface before the feature ($k < \sigma k_T$), the corrections appear as ascending powers of $k/k_T$, physically encoding how lingering spatial gradients scatter off the subsequent sudden shift in background kinetic energy. Conversely, for modes crossing after the transition ($k > \sigma k_T$), the corrections scale as powers of $k_T/k$, explicitly mirroring the decay rate of the transient decaying-mode component excited by the sharp transition. Standard $\delta N$ tracks only the growing mode and drops $k$-dependent spatial interactions entirely, leading to inherent miscalculations of the ringing phase and amplitude. By explicitly restoring both the spatial gradient interactions and the transition-induced decaying modes, the gradient-corrected framework interpolates smoothly between regimes and restores the accurate wave geometry similar to the full MS solution.

\subsubsection{Jacobian with Full Gradient Interactions}
\label{Sec_J_Full_Gradient}

We have shown that by employing the Green's function method within the CSA framework, the evaluation of all integrals for the $\mathcal{O}(k^2)$ gradient corrections reduces to straightforward polynomial integration. Consequently, this allows for a fully analytic evaluation of both the adiabatic and non-adiabatic mode corrections. However, extending this method to evaluate all gradient corrections becomes highly impractical. In this section, we use a simple identification that allows for the evaluation of the $\delta N$ power spectrum with full gradient interactions, circumventing the need to compute each order of the gradient expansion individually.

Since we are employing the basis $X = (\mathcal{R}_*, \mathcal{R}'_*)$, the linear $\delta N$ expansion takes the form
\begin{equation}
	\delta N = - \frac{1}{\Pi} \left( J^1_1 \mathcal{R}_* + J^1_2 \mathcal{R}'_* \right).
\end{equation}
A comparison with the relation $\delta N = - \hat{\mathcal{R}} = -\mathcal{R}_* u_{\rm ad} - \mathcal{R}'_* u_{\rm nad}$ suggests that if the parameter $\hat{\mathcal{R}}$ is constructed by retaining all spatial gradients, the following identifications must hold:
\begin{equation}
	C_1 = u_{\rm ad} = \frac{J^1_1}{\Pi}, \qquad
	C_2 = u_{\rm nad} = \frac{J^1_2}{\Pi}.
	\label{identification_full}
\end{equation}
Strictly speaking, this identification is valid only in the small $\epsilon$, according to the discussion in Section \ref{Equivalence_CSA_MS}, unless the momentum-corrected source term \eqref{corrected_source_term} is being employed. Substituting $C_a$ into the sensitivity equations~\eqref{SO_J_evolution} yields
\begin{equation}
	(J^1_a)'' + 2 a H (J^1_a)' + a^2 V_{,\phi\phi} J^1_a + k^2 J^1_a = 0,
	\label{J_full_gradient}
\end{equation}
which we expect to govern the full Jacobian evolution encompassing full spatial gradients. 

Note that in the $\delta N$ evaluations discussed previously, the forcing matrix $\Sigma^2_a$ relies on an $\hat{\mathcal{R}}$ derived via a piecewise matching procedure. This necessitated separating the evolution of the Jacobians for modes that reach the matching surface either before or after the transition. Here, however, because $\hat{\mathcal{R}}$ does not appear in the formalism, no such piecewise matching is required. The resulting $\delta N$ components simply evolve identically to the exact MS solution, automatically capturing the sub-horizon quantum phase oscillations and Bogoliubov mixing typically omitted by the standard SUA. Therefore, matching Jacobians to the MS solution at an arbitrary reference time sufficiently deep inside the horizon (imposing standard Bunch-Davies initial conditions) yields a globally accurate power spectrum.

However, in the subsequent section, where we evaluate the evolution of the Hessians relevant for non-Gaussianities, it is not possible to determine their evolution analytically while retaining the full gradient correction. Consequently, the Jacobians appearing in the second-order sensitivity equation \eqref{eq:Hessian_system} must still be evaluated using the same piecewise matching method applied to the Hessians. Thus, to maintain consistency for the later Hessian evaluation, we will still solve the linear sensitivities at the explicit horizon-crossing matching time $\sigma a H = k$.

Let us first consider the case of matching before the transition, $k < \sigma k_T$. For the piecewise linear potential of the Starobinsky model, relation \eqref{J_full_gradient} reduces to
\begin{equation}
	(J^1_a)'' - \frac{2}{\tau} (J^1_a)' + k^2 J^1_a = 0.
\end{equation}
The mode solutions for the two stages of the model take the form
\begin{equation}
	J^1_{a\pm} = \alpha_{a\pm} f(k \tau) + \beta_{a\pm} g(k \tau),
	\label{full_gradient_general_solution}
\end{equation}
where for notational simplicity, we have defined the functions
\begin{equation}
	f(x) = \sin (x) - x \cos (x), \qquad
	g(x) = \cos (x) + x \sin (x).
\end{equation}
Applying the initial conditions \eqref{Js_tau}, the integration constants are uniquely determined as
\begin{subequations}
\begin{align}
	\alpha_{1+} &= -\frac{A_+}{3 H^2 \sigma} \cos (\sigma), & \beta_{1+} &= -\frac{A_+}{3 H^2 \sigma} \sin (\sigma),  \\
	\alpha_{2+} &= -\frac{A_+}{3 H^2 k \sigma^2} g(\sigma), & \beta_{2+} &=  -\frac{A_+}{3 H^2 k \sigma^2} f(\sigma).
	\label{Full_gradient_Coefficients}
\end{align}
\end{subequations}
For the post-transition period, the jump conditions must be applied to properly transmit the Bogoliubov particle production generated by the sudden feature. Notice that the boundary conditions \eqref{boundary_J_star}, which were evaluated for the homogeneous case, also apply here. Using these for the integration constants in the post-transition period, we find

\begin{subequations}
	\begin{align}
		\alpha_{1-} &= -\frac{A_+}{3 H^2 \sigma } \Big[ \frac{3 \Delta A}{A_+ \tilde{k}^3} \tilde{g} \mathcal{U} + \cos(\sigma) \Big], 
		&\beta_{1-} &= -\frac{A_+}{3 H^2 \sigma } \Big[ \frac{3 \Delta A}{A_+ \tilde{k}^3} \tilde{f} \mathcal{U} +  \sin(\sigma) \Big], \\
		\alpha_{2-} &= -\frac{A_+}{3 H^2 k \sigma } \Big[ \frac{3 \Delta A}{A_+ \tilde{k}^3} \tilde{g} \mathcal{V} +  g(\sigma) \Big], 
		&\beta_{2-} &= -\frac{A_+}{3 H^2 k \sigma } \Big[ \frac{3 \Delta A}{A_+ \tilde{k}^3} \tilde{f} \mathcal{V} +  f(\sigma) \Big],
	\end{align}
\end{subequations}
where the notation $\tilde{f} \equiv f(\tilde{k})$ and $\tilde{g} \equiv g(\tilde{k})$ is used for conciseness, and we have introduced the functions
\begin{equation}
	\mathcal{U} \equiv \sin(\sigma) \tilde{g} - \cos(\sigma) \tilde{f}, \qquad
	\mathcal{V} \equiv f(\sigma) \tilde{g} - g(\sigma) \tilde{f}.
\end{equation}
Taking the superhorizon limit $\tau_e \to 0$, we obtain the full-gradient components evaluated at the end of inflation:
\begin{subequations}\label{J1a_full_gradients}
	\begin{align}
		\frac{J^1_1(\tau_e)}{\Pi_e}  &= \frac{A_+}{A_- \sigma } \Big[ \frac{3 \Delta A}{A_+ \tilde{k}^3} \tilde{f} \mathcal{U} +  \sin(\sigma) \Big], \\
		\frac{J^1_2(\tau_e)}{\Pi_e}  &= \frac{A_+}{A_- k \sigma ^2} \Big[ \frac{3 \Delta A}{A_+ \tilde{k}^3} \tilde{f} \mathcal{V} +  f(\sigma) \Big].
	\end{align}
\end{subequations}

The complete information regarding spatial gradient interactions is encoded within relations \eqref{J1a_full_gradients}. To verify this, we perform a Taylor expansion of these solutions in the long-wavelength limit to find:
\begin{subequations}\label{k4_corrections}
	\begin{align}
		\frac{J^1_1(\tau_e)}{\Pi_e}  &= 1 - \frac{\Delta A \tilde{k}^3}{3 A_- \sigma} + \frac{2 \Delta A \tilde{k}^2}{5 A_- } - \frac{\sigma^2}{6} \nonumber \\
		&\quad + \frac{\Delta A \tilde{k}^5}{15 A_-  \sigma} - \frac{6 \Delta A \tilde{k}^4}{35 A_-} + \frac{\Delta A \tilde{k}^3 \sigma}{6 A_-} - \frac{\Delta A \tilde{k}^2 \sigma^2}{15 A_- } + \frac{\sigma^4}{120}, \\
		\frac{J^1_2(\tau_e)}{\Pi_e}  &= \frac{\sigma}{3 k} \bigg[ 1 - \frac{\Delta A \tilde{k}^3}{A_- \sigma^3} + \frac{\Delta A \tilde{k}^5}{5 A_-  \sigma^3} - \frac{\Delta A \tilde{k}^3}{2 A_-  \sigma} + \frac{2 \Delta A \tilde{k}^2}{5 A_- } - \frac{\sigma^2}{10} \nonumber \\
		&\quad - \frac{3 \Delta A \tilde{k}^7}{175 A_- \sigma^3} + \frac{\Delta A \tilde{k}^5}{10 A_-\sigma} - \frac{6 \Delta A \tilde{k}^4}{35 A_- } + \frac{\Delta A \tilde{k}^3 \sigma}{8 A_- } - \frac{\Delta A \tilde{k}^2 \sigma^2}{25 A_-} + \frac{\sigma^4}{280} \bigg],
	\end{align}
\end{subequations}
where we have used the fact that the matching parameter $\sigma$ scales proportionally with $k$. A direct comparison with relations \eqref{GC_J_1} reveals that the first line of each expression above correctly reproduces the complete $\mathcal{O}(k^2)$ corrections derived previously. The subsequent lines detail the higher-order $\mathcal{O}(k^4)$ corrections. Thus, one can expand the full gradient solution to any desired order to isolate the respective spatial interactions.

For matching after the transition, $k > \sigma k_T$, the evolution of modes lies entirely on the $A_-$ branch. While the spatial mode equation formally retains its unperturbed de Sitter structure, the background evolution inherently retains the memory of the transition through a relaxation tail. Imposing the initial conditions $J_*$ onto $J^1_{a-}$, we obtain
\begin{subequations}
\begin{align}
	\alpha_{1-} &= \frac{-\Delta A}{3 H^2 \tilde{k}^3} \left[\cos (\sigma ) \left(\frac{A_- \tilde{k}^3}{\Delta A \sigma }-\sigma ^2\right)+3 g(\sigma )\right],
	&\alpha_{2-} &= \frac{-A_-}{3 H^2 k} \left[\frac{1}{\sigma ^2}-\frac{\Delta A \sigma }{A_- \tilde{k}^3}\right] g(\sigma ),\\
	\beta_{1-} &= \frac{-\Delta A}{3 H^2 \tilde{k}^3} \left[\sin (\sigma ) \left(\frac{A_- \tilde{k}^3}{\Delta A  \sigma }-\sigma ^2\right)+3 f(\sigma )\right],
	&\beta_{2-} &= \frac{-A_-}{3 H^2 k} \left[\frac{1}{\sigma ^2}-\frac{\Delta A \sigma }{A_- \tilde{k}^3}\right] f(\sigma ).
\end{align}
\end{subequations}
Mapping the components to the end of inflation $\tau_e \to 0$ yields:
\begin{subequations}\label{J1a_full_gradients2}
	\begin{align}
		\frac{J^1_1(\tau_e)}{\Pi_e}  &= \frac{\Delta A }{A_- \tilde{k}^3} \left[\sin (\sigma ) \left(\frac{A_- \tilde{k}^3}{\Delta A \sigma }-\sigma ^2\right)+3 f(\sigma )\right] , \\
		\frac{J^1_2(\tau_e)}{\Pi_e}  &= \frac{1}{k} \left[\frac{1}{\sigma ^2}-\frac{\Delta A \sigma }{A_- \tilde{k}^3}\right] f(\sigma ).
	\end{align}
\end{subequations}
We can similarly perform a long-wavelength expansion of these solutions to extract the post-transition gradient corrections up to $\mathcal{O}(k^4)$:
\begin{subequations}\label{k4_corrections2}
	\begin{align}
		\frac{J^1_1(\tau_e)}{\Pi_e}  &= 1 + \frac{\Delta A  \sigma^5}{15 A_- \tilde{k}^3} - \frac{\sigma^2}{6}  - \frac{\Delta A \sigma^7}{210 A_- \tilde{k}^3} + \frac{\sigma^4}{120}, \\
		\frac{J^1_2(\tau_e)}{\Pi_e}  &= \frac{\sigma}{3 k} \bigg[ 1 - \frac{\Delta A \sigma^3}{A_- \tilde{k}^3} + \frac{\Delta A  \sigma^5}{10 A_- \tilde{k}^3} - \frac{\sigma^2}{10}  - \frac{\Delta A \sigma^7}{280 A_- \tilde{k}^3} + \frac{\sigma^4}{280} \bigg],
	\end{align}
\end{subequations}
which, again, are consistent with the leading-order corrections derived in \eqref{GC_J_2} using the Green's function method.

The analytic $\delta N$ power spectrum, augmented with spatial corrections up to $\mathcal{O}(k^4)$ via \eqref{k4_corrections} and \eqref{k4_corrections2}, as well as the power spectrum constructed from the full-gradient solutions \eqref{J1a_full_gradients} and \eqref{J1a_full_gradients2}, are plotted in Figure \ref{powespectrum_starobinsky_model}. It is evident that the full-gradient case perfectly matches the MS solution. This equivalence can also be demonstrated analytically. By utilizing the pre-transition part of the solution in equation \eqref{pre_transition_nu} to determine $\mathcal{R}_*$ and $\mathcal{R}'_*$, the explicit form of the linear $\delta N$ for $k < \sigma k_T$ can be expressed as
\begin{align}
	-\delta N &= \frac{1}{\Pi_e} \left(J^1_{1+}(\tau_e) \mathcal{R}_* + J^1_{2+}(\tau_e) \mathcal{R}'_*\right) \nonumber \\
	&= \frac{3 H^3}{2 \sqrt{2} A_- A_+ k^{9/2}} \left[ -2 i A_+ k^3 + 3 k_T \Delta A \left( k^2 + k_T^2 + e^{\frac{2 i k}{k_T}} (k + i k_T)^2 \right) \right].
	\label{dN_equal_R}
\end{align}
Similarly, by using the post-transition part of the solution \eqref{pre_transition_nu}, an analogous relation for $\delta N$ can be derived for $k > \sigma k_T$. This $\sigma$-independent result is precisely the solution to the MS equation, as given in equation \eqref{R_star}. This confirms that the full-gradient $\delta N$ formalism successfully reproduces the MS solution, irrespective of the choice of matching time.

\subsection{Second-Order Sensitivities: Homogeneous Hessian}
\label{SO_CSA_Staro}


We now proceed to perform a second‑order CSA to analytically obtain the gradient‑corrected non‑Gaussianity \cite{komatsu2001acoustic, maldacena2003non, bartolo2004non, liguori2010primordial, chen2010primordial, wang2014inflation, cai2018revisiting, namjoo2025geometry}. In the homogeneous and small $\epsilon$ limit, the sensitivity equations for the Hessian \eqref{eq:Hessian_system} simplify as
\begin{equation}
	(\Theta^1_{ab})'' + 2 a H (\Theta^1_{ab})' + a^2 V_{,\phi\phi} \Theta^1_{ab} + a^2 V_{,\phi\phi\phi} J^1_a J^1_b= 0.
	\label{eq_Hessian_final_ODE}
\end{equation}
Away from the transition in the linear Starobinsky model, the second and third derivatives of the potential vanish. As a result, in both stages of the model, the above relation reduces to
\begin{equation}
	(\bar{\Theta}^1_{ab})'' -\frac{2}{\tau} (\bar{\Theta}^1_{ab})'= 0 \,.
\end{equation}
This differential equation admits the exact solution
\begin{equation}
	\bar{\Theta}^1_{ab\pm} = \gamma_{ab\pm} \tau^3 + \theta_{ab\pm} \,,
\end{equation}
where $\gamma_{ab\pm}$ and $\theta_{ab\pm}$ are stage-dependent integration constants.

As discussed in Section \ref{dn_homogeneous} regarding the linear $\delta N$ formalism, the discontinuity in the slope of the potential manifests as a Dirac delta function in the second derivative $V_{,\phi \phi}$. This, in turn, induces a discontinuity in the first derivative of the Jacobian, $(J^1_a)'$. When extending this analysis to the second-order sensitivity equation \eqref{eq_Hessian_final_ODE}, we encounter both the second and third derivatives of the potential. Specifically, the third derivative takes the form
\begin{equation}
	V_{,\phi\phi\phi} = -\frac{\Delta A}{|\phi'_T| \phi'(\tau)} \delta'(\tau - \tau_T) \,.
\end{equation}
The presence of the derivative of the Dirac delta function, $\delta'(\tau - \tau_T)$, is highly consequential: it induces not only a jump in the derivative of the Hessian but also a discontinuity in the Hessian itself. Therefore, a careful treatment of the matching conditions at the transition surface is required. 

To determine the boundary condition for the jump of the Hessian, $\Delta \Theta^1_{ab} \equiv \Theta^1_{ab}(\tau_T^+) - \Theta^1_{ab}(\tau_T^-)$, we isolate the most singular term in the sensitivity equation. Because the $\delta'$ term rigorously dominates the integral across the transition, it suffices to approximate the equation in the immediate vicinity of $\tau_T$ as
\begin{equation}
	(\Theta^1_{ab})'' = - a^2 V_{,\phi\phi\phi} J^1_a J^1_b = \frac{a^2 \Delta A}{|\phi'_T| \phi'(\tau)} J^1_a J^1_b \delta'(\tau - \tau_T).
\end{equation}
Integrating this expression once across the transition yields a Dirac delta function in the first derivative, $(\Theta^1_{ab})'$, and integrating a second time evaluates the singular distribution to yield a Heaviside step function in the background, representing a finite jump in the Hessian:
\begin{equation}
	\Delta \Theta^1_{ab} = \frac{a_T^2 \Delta A}{|\phi'_T| \phi'_T} J^1_a(\tau_T) J^1_b(\tau_T)\,.
	\label{boundary_Hessian}
\end{equation}

Next, we must determine the boundary condition for the derivative of the Hessian. Because $\Theta^1_{ab}$ undergoes a discontinuous jump at $\tau_T$, its first order derivative must contain a singular component proportional to a Dirac delta function alongside its regular continuous evolution. Therefore, at the transition, it takes the form:
\begin{equation}
	(\Theta^1_{ab})'= (\Theta^1_{ab})'_{reg} + \Delta \Theta^1_{ab} \delta (\tau - \tau_T) \,.
\end{equation}
To find the matching condition for the derivative jump, $\Delta (\Theta^1_{ab})'$, we integrate the full homogeneous part of the sensitivity equation \eqref{eq_Hessian_final_ODE} across the infinitesimal interval $[\tau_T^-, \tau_T^+]$. Handling the cross-terms between the Heaviside step functions and Dirac delta functions via integration by parts yields
\begin{align}
	\Delta (\Theta^1_{ab})' = \left.\left[\frac{2}{\tau} \Delta \Theta^1_{ab}  + \frac{a^2 \Delta A}{|\phi'| \phi'} \left(\Theta^1_{ab} \phi'+ \frac{1}{\tau} \left(1-\frac{\eta}{2}\right) J^1_a J^1_b- (J^1_a J^1_b)'\right)\right]\right|_{\tau_T}.
	\label{boundary_derivative_hessian}
\end{align}
Notice that the right-hand side of this expression involves $\Theta^1_{ab}$, $(J^1_a)'$, and the Hubble flow parameter $\eta$, all evaluated at the transition time $\tau_T$. Because these quantities are discontinuous at this boundary, standard distribution theory dictates that they must be evaluated using their symmetric average across the transition. Expressing these in terms of their pre-transition limits, we have:
\begin{subequations}
	\begin{align}
		\Theta^1_{ab}(\tau_T) &= \frac{1}{2} \left(\Theta^1_{ab}(\tau^+_T) + \Theta^1_{ab}(\tau^-_T)\right) = \Theta^1_{ab+}(\tau_T) + \frac{1}{2} \Delta \Theta^1_{ab},\\
		(J^1_{a})'(\tau_T) &= \frac{1}{2} \left((J^1_{a})'(\tau^+_T) + (J^1_{a})'(\tau^-_T)\right) = (J^1_{a+})'(\tau_T) + \frac{1}{2} \Delta (J^1_{a})', \\
		\eta(\tau_T) &= \frac{1}{2} \left(\eta(\tau^+_T) + \eta(\tau^-_T)\right) = \frac{3 \Delta A}{A_+}.
	\end{align}
\end{subequations}
Substituting into relation \eqref{boundary_derivative_hessian} and making use of equations \eqref{boundary_J_star} and \eqref{boundary_Hessian}, we obtain the following simplified boundary condition:
\begin{equation}
	\Delta (\Theta^1_{ab})' = \left. \frac{9 \Delta A H^2}{A_+^2} \left[\frac{A_+ k_T}{3 H^2} \Theta^1_{ab+}  + 3 k_T J^1_{a} J^1_{b}+ \left(J^1_{a+} J^1_{b+} \right)'\right]\right|_{\tau_T}.
	\label{boundary_Hessian_p}
\end{equation}

Having the boundary conditions specified, we can now find the evolution of the homogeneous Hessian. As before, we begin with the case of matching before the transition, $k < \sigma k_T$. Using the initial condition $\Theta^1_{ab}(\tau_*) = 0$, we see that all components of the Hessian vanish identically in the pre-transition period, i.e., $\bar{\Theta}^1_{ab+}(\tau) = 0$. For the post-transition evolution, we employ the boundary conditions \eqref{boundary_Hessian} and \eqref{boundary_Hessian_p} to obtain
\begin{subequations} \label{gamma_theta_general}
\begin{align}
	\gamma_{ab-} &=\frac{k_T^2}{3} \Delta( \Theta^1_{ab} )'  =\frac{3 \Delta A H^2 k_T^2}{A_+^2} \left.\left[3 k_T J^1_{a} J^1_{b}+ \left(J^1_{a+} J^1_{b+} \right)'\right]\right|_{\tau_T},\\
	\theta_{ab-} &= \Delta \Theta^1_{ab} + \frac{1}{3 k_T} \Delta( \Theta^1_{ab} )' = \left.\frac{3 \Delta A H^2}{A_+^2 k_T} \left[\left(J^1_{a+} J^1_{b+} \right)'\right] \right|_{\tau_T}.
\end{align}
\end{subequations}
Using the homogeneous Jacobians \eqref{HMJ_solution_staro}, we determine the integration constants for the Hessian to be
\begin{subequations}
	\begin{align}
		\gamma_{11-} &= \frac{\Delta A k_T^3}{H^2},
		&\theta_{11-} &= 0, \\
		\gamma_{12-} &= \gamma_{21-} = \frac{\Delta A k_T^2 \sigma}{3 H^2 \tilde{k}},
		&\theta_{12-} &= \theta_{21-} = \frac{\Delta A \tilde{k}^2}{3 H^2 k_T \sigma^2}, \\
		\gamma_{22-} &= \frac{\Delta A k_T(\sigma^6 - \tilde{k}^6)}{9 H^2 \tilde{k}^2 \sigma^4},
		&\theta_{22-} &= \frac{2 \Delta A \tilde{k} (\sigma^3 - \tilde{k}^3)}{9 H^2 k_T^2 \sigma^4}.
	\end{align}
\end{subequations}

Consequently, the Hessian evaluated at the end of inflation is given by $\bar{\Theta}^1_{ab-} (\tau_e) \approx \theta_{ab-}$. To evaluate the matrix $M_{ab}$ defined in equation \eqref{M_for_fNL}, it is sufficient to note that at late times $\eta$ and $J^2_a$ vanish. Therefore, $M_{ab} \approx \bar{\Theta}^{1}_{ab-}(\tau_e) \approx \theta_{ab-}$. Finally, to evaluate the non-Gaussianity parameter, we must also determine the power spectra required for formula \eqref{fNL_eq_CSA}. Using the pre-transition relation for the curvature perturbation \eqref{pre_transition_nu}, we readily obtain:
\begin{equation}
	\mathcal{P}^{11} = \frac{9 H^6 \left(\sigma^2+1\right)}{4 \pi^2 A_+^2}, \qquad
	\mathcal{P}^{12} = \mathcal{P}^{21} = -\frac{9 H^6 k \sigma}{4 \pi^2 A_+^2}, \qquad
	\mathcal{P}^{22} = \frac{9 H^6 k^2 \sigma^2}{4 \pi^2 A_+^2}.
	\label{power_ab_post}
\end{equation}
Substituting the evaluated matrix $M_{ab}$ and the power spectra $\mathcal{P}^{ab}$ into equation \eqref{fNL_eq_CSA}, the analytical expression for the equilateral-type non-Gaussianity is found to be
\begin{equation}
	f_{\rm NL}^{\rm eq} = -5 r K \frac{ \left[r K-\sigma ^2+3\right] \left[\sigma ^2 (3 -r K-K)+(K+3) (r K+3)+\sigma ^4\right] }{ \left[\sigma ^2 (3-2 r K)+(r K+3)^2+\sigma ^4\right]^2 },
	\label{fNL_staro}
\end{equation}
where we have defined the dimensionless parameters
\begin{equation}
	r \equiv \frac{\Delta A}{A_-}, \qquad K \equiv \frac{\tilde{k}^3}{\sigma}.
\end{equation}

For modes matching after the transition, $k>\sigma k_T$, applying the initial condition $\Theta^1_{ab}(\tau_*) = 0$ to the post-transition Hessian, yields $\bar{\Theta}^1_{ab-}(\tau_*) = 0$. Consequently, the first term in the matrix $M_{ab}$, defined in equation \eqref{M_for_fNL}, vanishes. The second and third terms also vanish due to the presence of $J^2_a$ and $\eta$, respectively. Therefore, $M_{ab}$ identically vanishes, and consequently, the resulting non-Gaussianity is zero. Obtaining a pure zero indicates that the non-Gaussianity in the post-transition period is of order $\mathcal{O}(\epsilon)$. Since we have neglected $\epsilon$ entirely in our analytical calculations, this vanishing result is expected.

\begin{figure}
	\centering
	\includegraphics[width=0.65\textwidth]{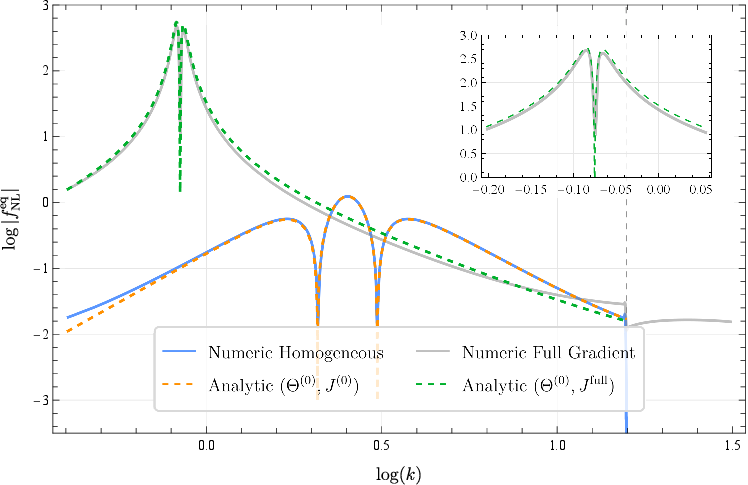}
	\caption{The logarithm of the absolute value of the equilateral-type non-Gaussianity parameter, $\log |f_{\rm NL}^{\rm eq}|$, for the Starobinsky model. The numerical values of the model parameters are chosen to be identical to those used in Figure \ref{powespectrum_starobinsky_model}. These parameters are adopted from Ref.~\cite{hazra2013bingo} to facilitate a direct comparison between our results and those obtained using the standard in-in formalism. The gray line represents the exact numerical integration of second-order CSA system with the momentum-corrected source term \eqref{corrected_source_term}, while the solid blue line shows the purely homogeneous numerical limit. These are compared against two analytic approximations: a purely homogeneous expansion $(\Theta^{(0)}, J^{(0)})$ shown by the orange dashed line \eqref{fNL_staro}, and a mixed approximation using the homogeneous Hessian and full-gradients Jacobian $(\Theta^{(0)}, J^{\text{full}})$ shown by the green dashed line \eqref{fNL_mix}. The vertical gray dashed line indicates the transition scale $k_T$, and the inset provides a magnified view of the gradient-corrected results close to the peak of the non-Gaussianity.}
	\label{fNL_starobinsky_model}
\end{figure}

The analytical expression for the $f_{\rm NL}^{\rm eq}$ parameter \eqref{fNL_staro} is illustrated in Figure \ref{fNL_starobinsky_model}, where it is compared with the full numerical homogeneous results. As seen in the figure, the analytical relation for the homogeneous $f_{\rm NL}^{\rm eq}$ accurately matches its numerical counterpart, thereby validating the correctness of our derivations. 

It is worth noting that even when employing the homogeneous Hessian, one can partially incorporate gradient interactions through the use of gradient-corrected Jacobians. By substituting the full gradient Jacobians in place of the homogeneous ones in relations \eqref{boundary_Hessian} and \eqref{boundary_Hessian_p}, we can determine the post-transition integration constants using precise boundary conditions. These gradient-corrected Jacobians can also be applied to the $f_{\rm NL}^{\rm eq}$ formula \eqref{fNL_eq_CSA}. Using the full gradient Jacobians \eqref{full_gradient_general_solution} in the general relations \eqref{gamma_theta_general}, the Hessian integration constants become:
\begin{subequations}
	\begin{align}
		\gamma_{11-} &= \frac{\Delta A k_T^3}{3 H^2 \sigma ^2} \mathcal{U} \left(2 \tilde{k}^2 \sin (\tilde{k}-\sigma )+3 \mathcal{U}\right),
		\qquad \theta_{11-} = \frac{2 \Delta A \, \tilde{k}^2 }{3 H^2 \sigma^2} \, \mathcal{U} \sin(\tilde{k}-\sigma), \\
		\gamma_{22-} &= \frac{\Delta A k_T}{3 H^2 \tilde{k}^2 \sigma ^4} \mathcal{V} \left(2 \tilde{k}^2 \mathcal{Y}+3 \mathcal{V}\right),
		\qquad \qquad \,\,\,\,\, \theta_{22-} = \frac{2 \Delta A }{3 H^2 k_T^2 \sigma^4}\, \mathcal{Y} \mathcal{V}, \\
		\gamma_{12-} &= \gamma_{21-} = \frac{\Delta A k_T^2}{3 H^2 \tilde{k} \sigma ^3} \left(\tilde{k}^2 \mathcal{U} \mathcal{Y}+\tilde{k}^2 \mathcal{V} \sin (\tilde{k}-\sigma )+3 \mathcal{U} \mathcal{V}\right), \\
		\theta_{12-} &= \theta_{21-} = \frac{\Delta A \, \tilde{k} }{3 H^2 k_T \sigma^3} \big(\mathcal{U} \mathcal{Y} + \mathcal{V} \sin(\tilde{k}-\sigma)\big),
	\end{align}
\end{subequations}
where we have defined a new parameter as
\begin{equation}
	\mathcal{Y} \equiv \sin (\tilde{k}-\sigma )+\sigma  \cos (\tilde{k}-\sigma ).
\end{equation}
These can then be used to find the late-time values $M_{ab} \approx \bar{\Theta}^1_{ab-} (\tau_e) \approx \theta_{ab-}$. Using this, alongside the pre-transition power spectra \eqref{power_ab_post}, one can analytically derive the following expression for the non-Gaussianity parameter:
\begin{equation}
	f_{\rm NL}^{\rm eq} = \frac{5 A_- \Delta A \, \tilde{k}^8 \left[ 3 \Delta A \tilde{f}^2 \big(A_+ \tilde{k}^4 - 3 \Delta A (\tilde{k}^2+1)\big) - A_+ \tilde{k}^3 \tilde{g} \big(A_+ \tilde{k}^3 \cos(\tilde{k}) + 6 \Delta A \tilde{f}\big) \right]}{3 \left[ \big(A_+ \tilde{k}^3 + 3 \Delta A \tilde{f} \tilde{g}\big)^2 + 9 \Delta A^2 \tilde{f}^4 \right]^2}.
	\label{fNL_mix}
\end{equation}
The behavior of this parameter is illustrated in Figure~\ref{fNL_starobinsky_model}. We choose numerical values for the Starobinsky model similar to those used in Ref.~\cite{hazra2013bingo}, allowing our plots to be directly compared with their results obtained using the numerical in--in formalism. It is evident that incorporating the gradient-corrected Jacobian significantly refines the predictions and provides a marked improvement over the results obtained using the standard $\delta N$ formalism.

Note that the non-Gaussianity parameter \eqref{fNL_mix} is $\sigma$-independent. This is a direct consequence of the specific method we employed in its construction. In equation \eqref{dN_equal_R}, it was shown that the combination $J^1_a \delta X^a$ is independent of $\sigma$ when using the full-gradient Jacobian. The denominator of the $f_{\rm NL}^{\rm eq}$ formula \eqref{fNL_eq_CSA} is simply the square of this expression. Furthermore, while the numerator contains the Hessian, an examination of the integration constants \eqref{gamma_theta_general} reveals that $\Theta^1_{ab}$ is proportional to the square of the Jacobian. Consequently, the numerator is proportional to the fourth power of $J^1_a \delta X^a$. Therefore, the entire expression becomes completely independent of the matching time.

Finally, it is important to clarify the scope and limitations of our second-order $\delta N$ calculations. In this work, as well as in Ref.~\cite{ahmadi2026delta}, the equivalence between the gradient-corrected $\delta N$ formalism and standard perturbation theory has been demonstrated at the linear level through a variety of analytical and numerical tests. In particular, the gradient-interaction source term was constructed by comparing the equations of motion in the SUA with those obtained from linear perturbation theory. Because this source term is derived only at linear order, the gradient-corrected $\delta N$ formalism cannot be expected to reproduce the results of second-order perturbation theory exactly. Achieving full equivalence at second order would require extending the construction of the source term by systematically comparing the SUA and perturbative equations at second order and deriving the corresponding nonlinear corrections.

Nevertheless, our analysis shows that the linear gradient source term alone already plays a significant role in second-order predictions. For instance, the non-Gaussianity relation~\eqref{fNL_eq_CSA} clearly indicates that using a Jacobian that neglects the leading gradient corrections would substantially affect the predicted amplitude of non-Gaussianity. It is therefore natural to expect that incorporating the gradient corrections—even if derived only at linear order—should improve the accuracy of the results. This expectation is tested by analytically evaluating $f_{\rm NL}^{\rm eq}$ both with and without the gradient source term, with the results shown in Figure~\ref{fNL_starobinsky_model}. These results indicate that the proposed source term captures an important part of the relevant physics even beyond linear order. A complete treatment, however, would require deriving the nonlinear corrections to the source term and establishing exact equivalence with second-order perturbation theory, which we leave for future work.

%

\section{Summary and Conclusions}

In this work, we have introduced Continuous Sensitivity Analysis (CSA) into the gradient-corrected $\delta N$ formalism, establishing a systematic and highly efficient framework to streamline the analytical evaluation of cosmological observables. While recent theoretical advancements have successfully incorporated spatial gradient interactions into the $\delta N$ formalism—thereby resolving the well-known limitations associated with the separate universe assumption—the practical implementation of the $\delta N$ formalism remains exceedingly difficult. 

To overcome this computational bottleneck, we utilize the CSA method, a rigorous mathematical technique that directly tracks the sensitivity of a dynamical system's final state to variations in its initial conditions. We derive the first- and second-order sensitivity equations that govern the evolution of the field Jacobian and Hessian. By reformulating the sensitivity of the final number of $e$-folds with respect to initial conditions as a system of coupled first-order differential equations, we recast the $\delta N$ formalism into a framework that is both analytically transparent and numerically efficient. This formulation eliminates the need to first derive an explicit analytic expression for the total number of $e$-folds and subsequently differentiate it with respect to initial field values. Instead, field trajectories are treated as continuous functions of their initial conditions, allowing the spatial gradient terms that arise in the $\delta N$ formalism to be handled readily.

To demonstrate the practical utility of this framework, we applied the CSA-based method to the Starobinsky model, which provides a prime example of an inflationary scenario featuring a sharp transition into an ultra-slow-roll phase. Within the CSA formulation, we employed the Green's function technique to evaluate gradient interactions, demonstrating that the leading-order $\mathcal{O}(k^2)$ corrections can straightforwardly be obtained analytically. Furthermore, we introduced an appropriate identification scheme that enables the derivation of the power spectrum while consistently incorporating full gradient interactions. We then employed the second-order sensitivity equations to compute the non-Gaussianity parameter, successfully capturing a relevant subset of the gradient contributions. Comparisons with full numerical calculations demonstrate excellent agreement; the resulting analytical expressions accurately reproduce the characteristic $k$-dependence of both the power spectrum and the equilateral non-Gaussianity parameter, $f_{\rm NL}^{\rm eq}$, even when the matching procedure is performed at horizon crossing.

Beyond practical calculations, the CSA approach yields valuable theoretical insights and computational benefits. We have shown that CSA facilitates a direct comparison between the $\delta N$ output and standard linear perturbation theory. Through this comparison, we provided a more rigorous proof of their equivalence and, crucially, identified regimes where the standard $\delta N$ formalism approaches its theoretical limits due to the absence of the momentum constraint. Specifically, we demonstrated that the formalism breaks down in regimes characterized by a large Hubble flow parameter, $\epsilon$. This finding has significant implications for the study of punctuated inflation, prompting us to propose a corrected source term that mimics the effects of the momentum constraint. From a computational perspective, the CSA framework significantly improves numerical efficiency by reducing the second-order evolution problem for curvature perturbations to a system of first-order differential equations, thereby enhancing both numerical stability and execution speed.

In conclusion, the integration of CSA into the $\delta N$ formalism provides a powerful and robust toolkit for the early-universe cosmology community. As the search for primordial black holes and the characterization of the stochastic gravitational-wave background intensify, the ability to simplify the inclusion of spatial gradients becomes vital for accurately tracking the transition of modes from sub-horizon to super-horizon scales. Future work will focus on extending this sensitivity framework to multi-field scenarios and developing robust theoretical corrections for large-$\epsilon$ regimes. Moreover, the CSA method is naturally suited for the evaluation of the full probability density function of curvature perturbations, where the precise inclusion of gradient-corrected nonlinearities remains a key frontier of current research.

\acknowledgments
We acknowledge the financial support of the Research Council of the University of Tehran. We also thank Nahid Ahmadi for helpful discussions and comments.

\appendix

\section{\boldmath $\delta N$ Expansion}
\label{dN_expansion_appendix}

We assume the comoving gauge on the end-of-inflation hypersurface, such that $\delta\phi_e = 0$. We expand $\phi_e = \phi(X,n_e)$ to second order in the phase-space perturbations $\delta X^a$ and the $e$-fold shift $\delta N $ as
\begin{align}
	0 = \delta\phi_e &=
	\left. \frac{\partial\phi}{\partial X^a} \right|_{n_e}\delta X^a 
	+ \left.\frac{\partial\phi}{\partial n}\right|_{n_e} \delta N 
	+ \frac{1}{2}\left.\frac{\partial^2\phi}{\partial X^a\partial X^b}\right|_{n_e} \delta X^a\delta X^b \nonumber\\
	&+ \left. \frac{\partial^2\phi}{\partial X^a \partial n }\right|_{n_e} \delta X^a\,\delta N 
	+ \frac{1}{2} \left.\frac{\partial^2\phi}{\partial n^2}\right|_{n_e} \delta N ^2 = 0.
	\label{eq:deltaphi_expansion}
\end{align}
Or, in our notation:
\begin{equation}
	J^1_a\,\delta X^a + \Pi \,\delta N 
	+ \frac{1}{2}\,\Theta^1_{ab}\,\delta X^a\delta X^b 
	+ J^2_a\delta X^a \delta N 
	+ \frac{\Pi \eta}{4}\,\delta N^2 = 0.
	\label{eq_constraint_full}
\end{equation}
Here all $\delta X^a$ are evaluated at the matching time, while all other quantities are evaluated at the end of inflation. For notational simplicity, we have suppressed the subscript $e$.

At first order, the constraint \eqref{eq_constraint_full} yields
\begin{equation}
	\delta N^{(1)} = -\frac{1}{\Pi}\,J^1_a\,\delta X^a .
\end{equation}
Substituting this result into the second-order terms, we solve for the second-order contribution,
\begin{equation}
	\label{eq:deltaN2_raw}
	\delta N^{(2)} = -\frac{1}{\Pi}\left[
	\frac{1}{2}\,\Theta^1_{ab}
	- \frac{1}{\Pi} J^1_a J^2_b
	+ \frac{\eta}{4 \Pi}\, J^1_a J^1_b
	\right]\delta X^a\delta X^b .
\end{equation}
Summing the first- and second-order contributions, the final $\delta N$ expression becomes
\begin{equation}
	\delta N = -\frac{1}{\Pi} \left[ {J}^1_a \delta X^a + \frac{1}{2} {M}_{ab} \delta X^a \delta X^b \right],
\end{equation}
where
\begin{equation}
	{M}_{ab} = {\Theta}^1_{ab} - \frac{2}{\Pi} {J}^2_a {J}^1_b + \frac{\eta}{2 \Pi} {J}^1_a {J}^1_b.
\end{equation}

\section{\boldmath Exact Evolution of Comoving Curvature Perturbation in SUA}

In deriving the SUA relation \eqref{perturbed_BG}, which was subsequently compared with its counterpart derived from the full linear perturbation equation \eqref{perturbed_KG}, the momentum constraint from linear perturbation theory was utilized \cite{pattison2019stochastic}. Its use is justified as long as one works in a regime where the lapse function perturbation $A$ vanishes, and employing the momentum constraint—which provides information regarding the interactions between distinct FLRW patches—does not jeopardize the SUA. Nevertheless, if one wishes to work in the exact SUA, the momentum constraint must be dropped, relying solely on equations derived from the FLRW metric with vanishing spatial gradients. This is the primary aim of this appendix.

Here, we perturb the background equations with the source term, keeping only first-order terms in linear perturbation theory. We then use these results to determine the exact evolution equation governing $\mathcal{R}$. These results can subsequently be compared with the output of the CSA formalism to verify its validity. For the sake of comparison with other works, we also perform the same calculations for the $\delta N$ formalism with spatial curvature discussed in Ref.~\cite{artigas2025extended}. These calculations demonstrate that the CSA-derived results are highly consistent with exact direct calculations in linear perturbation theory, and that the formulation is flexible enough to be implemented in different $\delta N$ frameworks.

\subsection{$\delta N$ Formalism With a Source Term}
\label{sec_dN_source_SUA}

The gradient-corrected $\delta N$ formalism relies on the background equations \eqref{BG_equations_corrected}. To evaluate the linearized form of these equations, we identify the inhomogeneous Hubble parameter with $\theta / 3$, where $\theta \equiv n^\mu_{\,\,;\mu}$ is the expansion rate of the $t = \text{constant}$ hypersurfaces. Considering the perturbed FLRW metric,
\begin{equation}
	ds^{2} = -(1 + 2A)\, dt^{2}
	+ 2a \partial_{i} B \, dx^{i} dt
	+ a^{2} \left[ (1 - 2\psi)\delta_{ij} + 2 \partial_{i}\partial_{j} E \right] dx^{i} dx^{j},
\end{equation}
the expansion rate can be evaluated as \cite{pattison2019stochastic, malik2009cosmological}
\begin{equation}
	\theta = 3
	\left(
	H - H A - \dot{\psi} + \frac{1}{3a^2} \nabla^{2} \left( a^2 \dot{E} - a B\right)
	\right).
	\label{expansion_rate_full}
\end{equation}

Note that the corrected $\delta N$ formalism with the source term implies that the patches still evolve independently, each governed by a set of corrected background equations of the form~\eqref{BG_equations_corrected}. This means that the gradient source term is added to the background equation of each patch to mimic the effect of missing gradient interactions. This differs from linear perturbation theory, where the constraints are imposed as independent equations obeyed globally across all patches. In the source-term formalism, we have a field equation, and no additional constraints are imposed. Therefore, since the patches are still assumed to be separate, equation \eqref{expansion_rate_full} yields the variation of the Hubble parameter as $\delta H = - H A - \dot{\psi}$, similar to the SUA. Using this, together with the mappings $dt \rightarrow (1+A) dt$ and $\phi \rightarrow \phi + \delta \phi$, the perturbed background equations \eqref{BG_equations_corrected} up to first order are given by
\begin{subequations}\label{perturbed_BG_sourceterm}
	\begin{align}
		-2 A V - H^2 \delta \phi_{,n} \Pi  - 6 H^2 \psi_{,n} - \delta \phi V_{,\phi} &=  \mathcal{S}_{\rm F}, \label{perturbed_Friedmann_BG}\\
		H^2 \left[ \delta \phi_{,nn} - A_{,n} \Pi  - \frac{1}{2}  \delta \phi_{,n} \Pi^2 + 3  \delta \phi_{,n} - 3 \psi_{,n} \Pi + 2 A V_{,\phi} \right]+ \delta \phi V_{,\phi\phi} &= \mathcal{S}_{\rm KG},
	\end{align}
\end{subequations}
where $\mathcal{S}_{\rm F}$ and $\mathcal{S}_{\rm KG}$ are defined as the source terms added to the Friedmann and KG equations, respectively. In the case of our study, $\mathcal{S}_{\rm F}$ is zero and $\mathcal{S}_{\rm KG}$ is given by \eqref{source_term}. Eliminating the lapse function between equations \eqref{perturbed_BG_sourceterm}, one can find the following relation for the comoving curvature perturbation $\mathcal{R} = \psi + \frac{1}{\Pi} \delta \phi$:
\begin{equation}
	\mathcal{R}_{,nn} + \left(3 - \epsilon + \frac{3 \eta }{3 - \epsilon} \right) \mathcal{R}_{,n} + \frac{k^2 }{a^2 H^2} \hat{\mathcal{R}} = 0,
	\label{R_evolve_exact_source}
\end{equation}
which should be compared with the exact MS equation \eqref{exact_MS_N}. The comparison immediately reveals that the two equations are identical, except for an additional $\epsilon$ term appearing in \eqref{R_evolve_exact_source}. This additional term appears here because the momentum constraint is missing in the set of linearized equations \eqref{perturbed_BG_sourceterm}, as also discussed in the recent review paper~\cite{cruces2022review}. In particular, if we write equation \eqref{R_evolve_exact_source} in terms of cosmic time and use the definition $Q = \dot{\phi} \mathcal{R} / H$, in the small $k$ limit one finds
\begin{equation}
	\ddot{Q}
	+ 3H \left(1 + \frac{\epsilon \eta}{3(3-\epsilon)} \right)\dot{Q}
	+ H^{2}
	\left(
	-\frac{3}{2}\eta
	+ \frac{1}{2}\epsilon\eta
	- \frac{1}{4}\eta^{2}
	- \frac{1}{2}\eta\xi
	- \frac{\epsilon^{2}\eta^{2}}{2(3-\epsilon)}
	\right)
	Q = 0 ,
\end{equation}
which matches equation~(69) in Ref.~\cite{cruces2022review}.
This analysis shows that equation \eqref{R_evolve_exact_source}, which is the exact evolution equation of our gradient-corrected $\delta N$ formalism, is highly consistent with the CSA output \eqref{compare_CSA}. This validates the accuracy of the CSA-based $\delta N$ calculation.

\subsection{$\delta N$ Formalism With Spatial Curvature}
\label{sec_dN_sc_SUA}

It is also instructive to evaluate the exact evolution of $\mathcal{R}$ in the $\delta N$ formalism with spatial curvature introduced in Ref.~\cite{artigas2025extended} and compare the results against CSA-based calculations. The background equations for an FLRW metric with non-vanishing spatial curvature $\mathcal{K}$ are given by
\begin{subequations}\label{BG_K_equations}
	\begin{align}
		H^2 \left( 3 - \frac{\Pi^2}{2} \right)- V
		&=  -3 \frac{a_*^2}{a^2} \mathcal{K},\\
		H \frac{d}{dn}\!\left(H \Pi\right) + 3H^2 \Pi + V_{,\phi} &= 0.
	\end{align}
\end{subequations}
The linearized forms of these equations take the form of equations \eqref{perturbed_BG_sourceterm} with the source terms
\begin{equation}
	\mathcal{S}_{\rm F} = -3 \frac{a_*^2}{a^2} \mathcal{K}, \qquad \mathcal{S}_{\rm KG} = 0.
\end{equation}
Eliminating the lapse function between the two linearized background equations and using the background equations \eqref{BG_K_equations}, we find
\begin{equation}
	\mathcal{R}_{,nn} + \left(3  -\epsilon + \frac{3 \eta }{3 - \epsilon} \right) \mathcal{R}_{,n} + \left(\frac{3 \eta }{2 \epsilon - 6 } - 2\right)\frac{a_*^2\mathcal{K}}{a^2 H^2} = 0.
	\label{Exact_SC_R}
\end{equation}

As expected, the coefficients of $\mathcal{R}_{,n}$ and $\mathcal{R}_{,nn}$ are exactly the same as in relation \eqref{R_evolve_exact_source}, but the gradient-sourced terms differ. It can be shown that by identifying the spatial curvature as
$
a_*^2 \mathcal{K} = \frac{2}{3} k^2 \mathcal{R}_*
$,
the linearized Friedmann equation \eqref{perturbed_Friedmann_BG} takes exactly the form of the Hamiltonian constraint in full linear perturbation theory at the matching time \cite{artigas2025extended}. With this choice, relation \eqref{Exact_SC_R} becomes
\begin{equation}
	\mathcal{R}_{,nn} + \left(3  -\epsilon + \frac{3 \eta }{3 - \epsilon} \right) \mathcal{R}_{,n} + \left(\frac{\eta }{\epsilon -3}-\frac{4}{3}\right) \frac{k^2 \mathcal{R}_*}{a^2 H^2} = 0.
	\label{exact_R_SC}
\end{equation}
To retain the effect of the leading-order adiabatic correction, which is the primary purpose of this formulation, the last term in the above equation should take the form $k^2 \mathcal{R}_*/(a^2 H^2)$. However, the relation above deviates from this structure. Nevertheless, as pointed out in Ref.~\cite{artigas2025extended}, this discrepancy can be resolved by recognizing that part of the $k^2$ correction should be determined from linear perturbations and added to the final result. Specifically, when using the expressions derived in Ref.~\cite{artigas2025extended} (e.g., their equations (41) and (49)), where the $k^2$ correction of $\mathcal{R}$ obtained from linear perturbation theory is added to the curvature term, one finds that the MS equation is satisfied including at order $k^2$ (provided that $\epsilon$ remains small). This demonstrates that the spatial curvature approach remains compatible with the source-term formalism discussed in Appendix~\ref{sec_dN_source_SUA}.


We now use the CSA to analytically find the evolution of $\mathcal{R}$ to be compared with the exact relation \eqref{exact_R_SC}. Eliminating the Hubble parameter from the background equations \eqref{BG_K_equations} yields
\begin{equation}
	\Pi_{,n} + \left(3-\frac{1}{2} \Pi^2\right) \left(\frac{V_{,\phi} + \frac{  a_*^2 \mathcal{K} }{a^2} \Pi}{V-\frac{3 a_*^2 \mathcal{K} }{a^2}}+\Pi\right) = 0.
\end{equation}
Since $\mathcal{K}$ is first order and we are interested in results at linear order in perturbation theory, we can safely expand the above expression as
\begin{equation}
	\Pi_{,n} + \left(3-\frac{1}{2} \Pi^2\right) \left(\frac{V_{,\phi}}{V}+\Pi\right)  +\frac{a_*^2 \mathcal{K} }{a^2 H^2} \left(\frac{3 V_{,\phi}}{V}+\Pi\right) = 0.
\end{equation}
Because the background part is untouched, the stability matrix \eqref{A_matrix_tau} remains unchanged. The forcing matrix can then be evaluated as
\begin{equation}
	\left[\Sigma^i_a\right] =
	-\frac{2 k^2}{a^2 H^2} \left(g+\frac{\Pi }{3}\right)
	\begin{pmatrix}
		0 & 0 \\
		1 & 0
	\end{pmatrix}.
\end{equation}
Using these results, we find the following first-order sensitivity equations for the spatial curvature $\delta N$ formalism:
\begin{subequations}
	\begin{align}
		({J}^1_1)_{,nn} &= (\epsilon - 3)g_{,\phi} {J}^1_1 + \Big[ \Pi(g + \Pi) + (\epsilon - 3) \Big] ({J}^1_1)_{,n} -\frac{2 k^2}{a^2 H^2} \left(g+\frac{\Pi }{3}\right), \\
		({J}^1_2)_{,nn} &= (\epsilon - 3)g_{,\phi} {J}^1_2 + \Big[ \Pi(g + \Pi) + (\epsilon - 3) \Big] ({J}^1_2)_{,n} .
	\end{align}
\end{subequations}
Combining these equations yields
\begin{equation}
	({J}^1_a \delta X^a)_{,nn} = (\epsilon - 3)g_{,\phi} ({J}^1_a \delta X^a) + \Big[ \Pi(g + \Pi) + (\epsilon - 3) \Big] ({J}^1_a \delta X^a)_{,n} -\frac{2 k^2}{a^2 H^2} \left(g+\frac{\Pi }{3}\right) \mathcal{R}_*,
\end{equation}
which leads to the following final result after employing the $\delta N$ expansion \eqref{linear_dN_expansion_tau}:
\begin{equation}
	\delta N_{,nn} + \left(3  -\epsilon + \frac{3 \eta }{3 - \epsilon} \right) \delta N_{,n} - \left(\frac{\eta }{\epsilon -3}-\frac{4}{3}\right) \frac{k^2 \mathcal{R}_*}{a^2 H^2} = 0.
\end{equation}
This clearly demonstrates that $\delta N$ in this formalism evolves exactly as $- \mathcal{R}$ in equation \eqref{exact_R_SC}.

\section{\boldmath Failure of $\delta N$ Formalism at Large $\epsilon$ Regimes}
\label{appen_punctuated_inflation}

In Section~\ref{Equivalence_CSA_MS}, we utilized the CSA formulation to demonstrate that the predictions of the full-gradient $\delta N$ formalism match the solutions of the MS equation, strictly within the small-$\epsilon$ limit (as formally deduced from Equation~\eqref{compare_CSA}). This limitation arises because the momentum constraint is absent in the exact SUA \cite{kodama1998evolution, sasaki1998super, cruces2022review}. Although the introduced source term formalism can successfully predict the effects of gradient interactions, certain subtle dynamics originating from the momentum constraint remain missing from the formalism. To explicitly illustrate this shortcoming, we investigate the paradigm of punctuated inflation \cite{jain2008double, jain2010tensor, hazra2013bingo} in this appendix.

Punctuated inflation is characterized by the presence of an inflection point or a step in the scalar potential. As the inflaton field traverses this region, the background dynamics severely deviate from the standard slow-roll approximation; notably, the first Hubble flow parameter, $\epsilon$, can transiently exceed unity. Physically, the field experiences a brief but rapid acceleration down a steep potential drop, immediately followed by a flat plateau. During this sharp descent, the field velocity increases drastically, causing $\epsilon > 1$. Once the field reaches the plateau, the kinetic energy rapidly redshifts away, Hubble friction dominates once more, and the system gracefully re-enters a standard slow-roll phase. This brief period of highly accelerated, non-slow-roll dynamics typically imprints a characteristic signature on the primordial observables, often leading to an enhancement in the scalar power spectrum at specific characteristic scales.

A minimal and representative realization of punctuated inflation is provided by the following polynomial potential \cite{jain2008double, jain2010tensor, hazra2013bingo}:
\begin{equation}
	V(\phi) = \frac{1}{2} m^2 \phi^2 
	- \frac{\sqrt{2\lambda (p-1)}\, m}{p}\phi^p 
	+ \frac{\lambda}{4}\phi^{2(p-1)},
\end{equation}
where $m$ and $\lambda$ are constant parameters, and $p > 2$ is an integer. The coefficients in this potential are meticulously chosen such that the potential possesses an inflection point at $\phi = \phi_0$, where both the first and second derivatives of the potential vanish simultaneously. The precise location of this inflection point is given by
\begin{equation}
	\phi_0 =
	\left(\frac{2 m^2}{(p-1)\lambda}\right)^{\frac{1}{2(p-2)}}.
\end{equation}

\begin{figure}
	\centering
	\includegraphics[width=0.48\textwidth]{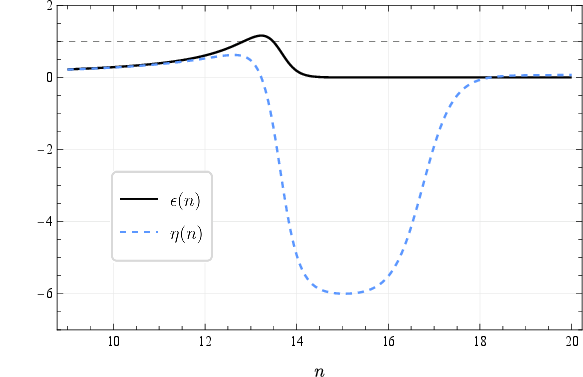}\,\,\,
	\includegraphics[width=0.48\textwidth]{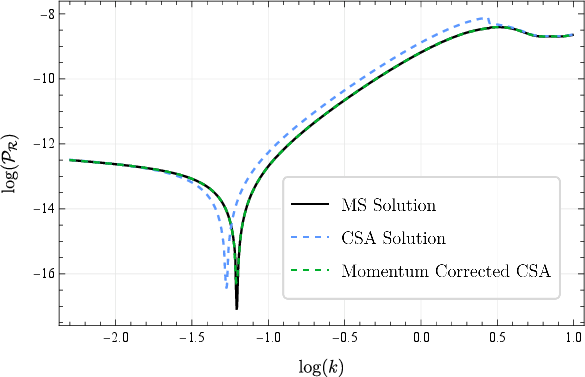}
	\caption{The left panel presents the first two Hubble flow parameters, while the right panel displays the power spectrum of curvature perturbations for the punctuated inflation model, obtained using a fully numerical approach. The power spectra are computed from the MS equation \eqref{MS_equation_R} (solid black line), the full-gradient $\delta N$ formalism with the source term \eqref{source_term} (dashed blue line), and the momentum-corrected source term \eqref{corrected_source_term} (dashed green line). The matching time is chosen at horizon crossing, $\sigma = 1$.
	}
	\label{Punctuated_inflation}
\end{figure}

For the specific case of $p=3$, the potential parameters that yield a phenomenologically viable primordial power spectrum are $m = 1.5012 \times 10^{-7}$ and $\phi_0 = 1.95964$. The inflaton field is assumed to start from rest at an initial value of $\phi_{\rm in} = 11.5$. The dynamical evolution of the first two Hubble flow parameters for this configuration is depicted in Figure~\ref{Punctuated_inflation}. As observed, $\epsilon$ briefly becomes larger than unity, making this model an ideal theoretical testbed for evaluating the robustness of the $\delta N$ formalism in large-$\epsilon$ regimes. 

We evaluated the gradient-corrected $\delta N$ power spectrum for this model using a fully numerical method, employing both the standard source term \eqref{source_term} and the momentum-corrected source term \eqref{corrected_source_term}. The results are illustrated in Figure~\ref{Punctuated_inflation}. It is immediately apparent that using the standard source term fails to accurately reproduce the exact MS solution during and after the transient non-slow-roll phase. However, the corrected source term provides an excellent fit to the exact MS solution. This outcome justifies the use of the source term~\eqref{corrected_source_term} throughout this work.

\bibliographystyle{JHEP}
\bibliography{CSA}

\end{document}